\documentclass[aps,pra,reprint,superscriptaddress,floatfix,longbibliography]{revtex4-2}
\usepackage[T1]{fontenc}
\usepackage[utf8]{inputenc}
\usepackage{lmodern}
\usepackage{amsmath,amssymb}
\usepackage[version=4]{mhchem}
\usepackage{graphicx}
\usepackage{microtype}
\usepackage{hyperref}
\usepackage{xurl}
\hypersetup{hidelinks,hypertexnames=false}
\begin{document}

\title{Solvent Fluctuations Enable Photoionization of Aqueous Nitrite Microdroplets}

\author{Brandon J. Wallace}
\thanks{These authors contributed equally to this work.}
\affiliation{Chemical Sciences Division, Lawrence Berkeley National Laboratory, Berkeley, California 94720, USA}
\affiliation{Department of Chemistry, University of California, Berkeley, California 94720, USA}

\author{Eric R. Heller}
\thanks{These authors contributed equally to this work.}
\affiliation{Chemical Sciences Division, Lawrence Berkeley National Laboratory, Berkeley, California 94720, USA}
\affiliation{Department of Chemistry, University of California, Berkeley, California 94720, USA}

\author{Alex Prophet}
\affiliation{Chemical Sciences Division, Lawrence Berkeley National Laboratory, Berkeley, California 94720, USA}

\author{Pranav Ballapuram}
\affiliation{Chemical Sciences Division, Lawrence Berkeley National Laboratory, Berkeley, California 94720, USA}
\affiliation{Department of Chemistry, University of California, Berkeley, California 94720, USA}

\author{Ashok Ajoy}
\affiliation{Chemical Sciences Division, Lawrence Berkeley National Laboratory, Berkeley, California 94720, USA}
\affiliation{Department of Chemistry, University of California, Berkeley, California 94720, USA}

\author{David T. Limmer}
\email{dlimmer@berkeley.edu}
\affiliation{Chemical Sciences Division, Lawrence Berkeley National Laboratory, Berkeley, California 94720, USA}
\affiliation{Department of Chemistry, University of California, Berkeley, California 94720, USA}
\affiliation{Materials Sciences Division, Lawrence Berkeley National Laboratory, Berkeley, California 94720, USA}
\affiliation{Kavli Energy NanoScience Institute, Berkeley, California 94720, USA}

\author{Kevin R. Wilson}
\email{krwilson@lbl.gov}
\affiliation{Chemical Sciences Division, Lawrence Berkeley National Laboratory, Berkeley, California 94720, USA}

\begin{abstract}
Nitrite photochemistry in atmospheric aerosols and natural waters is a major source of hydroxyl radicals and reactive nitrogen oxides. Contrary to the prevailing view that in the environment aqueous nitrite photochemistry proceeds exclusively through photoexcitation followed by homolytic bond cleavage, we show that visible light directly ionizes aqueous nitrite in individual micron-sized droplets. Although the photoionization cross section is exceptionally small, this pathway is observable at the high photon flux densities accessible in single-droplet experiments. Molecular simulations reveal that this is because photoionization absorption has an exceptionally long wavelength tail due to the large solvent reorganization energy in aqueous solutions. Kinetic modeling of an aqueous-phase reaction mechanism quantitatively accounts for the observed chemistry and predicts the formation of nitrate, hydroxyl radicals, and solvated electrons. The overlap of the predicted broad photoionization continuum with the visible solar spectrum further suggests that visible-light photoionization is an overlooked source of nitrogen oxides, solvated electrons, and oxidants in atmospheric aerosols, snowpack, and natural waters. Under dilute conditions, where dissolved oxygen is the main sink for solvated electrons, photoionization is expected to be the dominant photochemical loss pathway for nitrite.
\end{abstract}

\maketitle

Nitrite ($\mathrm{NO}_{2}^{-}$) and nitrate ($\mathrm{NO}_{3}^{-}$) are central to the global nitrogen cycle,\cite{ref2,ref3} where aqueous photochemistry converts these relatively stable ions into reactive nitrogen species and oxidants.\cite{ref4,ref5} Although aqueous nitrite photochemistry is conventionally attributed to electronic excitation followed by bond cleavage forming hydroxyl radicals ($\mathrm{OH}^{\bullet}$),\cite{ref1} we report evidence for a new photoionization pathway of aqueous nitrite that is operative at visible wavelengths. Although visible photoionization is a rare event, theoretical calculations provide evidence for an extended long wavelength tail that overlaps with a substantial portion of the visible solar spectrum, suggesting that photoionization could play a previously unrecognized role in controlling the lifetime of nitrite and provide a new mechanism for the formation of highly reactive species such as solvated electrons in natural waters.

Given its important role in environmental and atmospheric chemistry, the photochemistry of aqueous nitrite is well-documented.\cite{ref1,ref6,ref7,ref8} The $\pi\rightarrow\pi^{*}$ transition exhibits a strong maximum at 205 nm with a molar extinction coefficient ($\epsilon$) of 5500 $\mathrm{M}^{-1}\,\mathrm{cm}^{-1}$. A second weaker $n\rightarrow\pi^{*}$ band ($\epsilon$ = 22 $\mathrm{M}^{-1}\,\mathrm{cm}^{-1}$) is centered at 360 nm. Excitation at these wavelengths is sufficient to break N-O bonds forming nitric oxide ($\mathrm{NO}$) and hydroxyl radicals ($\mathrm{OH}^{\bullet}$, at pH < 12). Subsequent reactions of $\mathrm{NO}$ and $\mathrm{OH}^{\bullet}$ form $\mathrm{NO}_{3}^{-}$. In small droplets, we observe the photon driven oxidation of nitrite to nitrate at 532 nm, a wavelength that is more than 150 nm red-shifted from the known absorption bands of aqueous $\mathrm{NO}_{2}^{-}$.

\begin{figure*}[t]
\centering
\includegraphics[width=\textwidth]{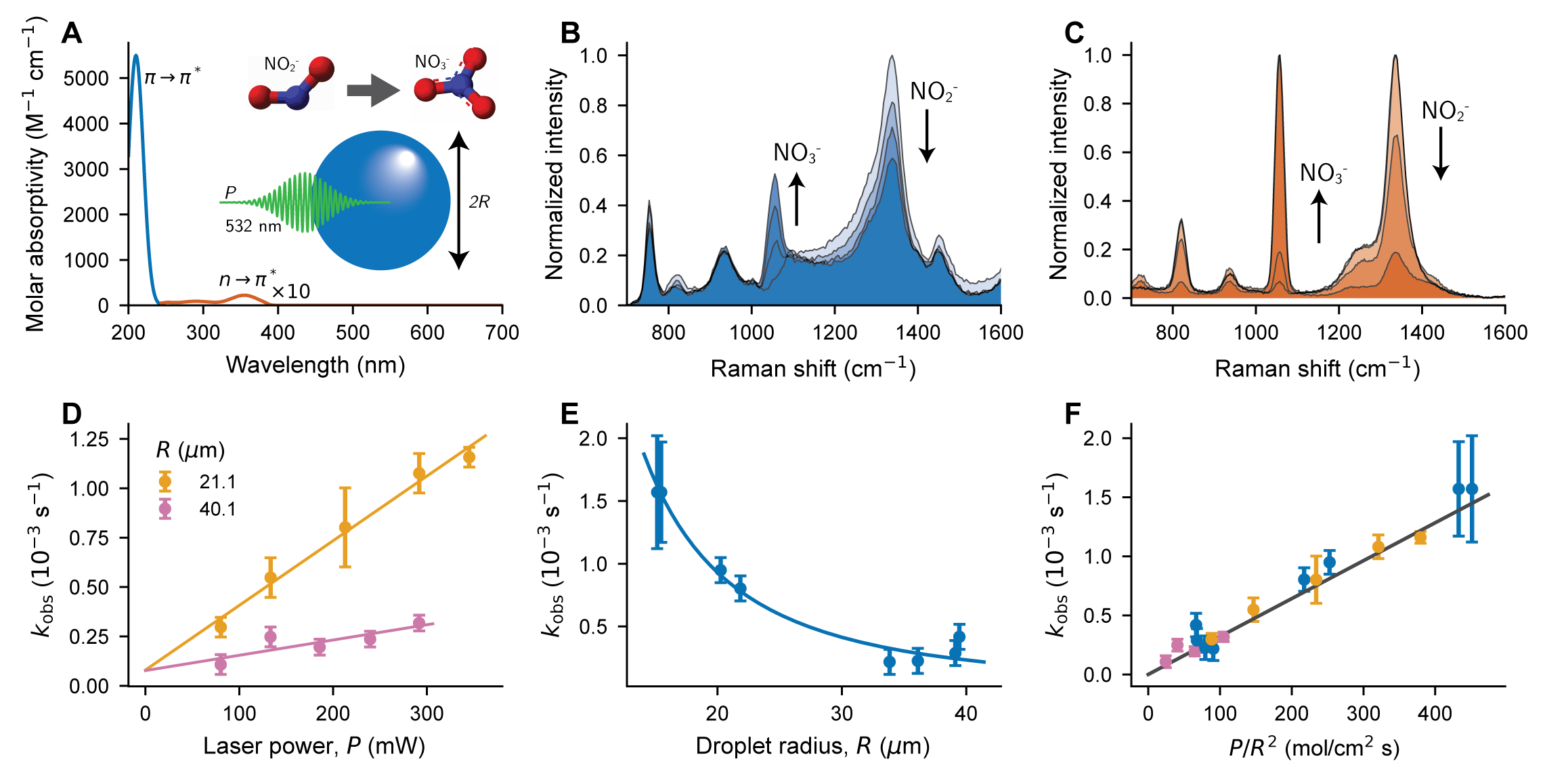}
\caption{Microdroplets irradiated with continuous wave 532 nm light convert nitrite to nitrate. (A) Absorption spectra\cite{ref1} showing $\pi\rightarrow\pi^{*}$ and $n\rightarrow\pi^{*}$ transitions for aqueous nitrite. Aqueous nitrite droplets with radius R formed in either a microfluidic chip or dispensed and levitated in a QET are illuminated with 532 nm light at power P. Raman spectra vs. reaction time for (B) microemulsions and (C) levitated droplets. The spectra of microemulsions have background spectral contributions from the oil phase and the microfluidic chip. (D) Laser dependent decay kinetics for microfluidic droplets initially containing 2.5~$\mathrm{M}$ nitrite for two different droplet sizes and corresponding linear fits. (E) Decay kinetics for initial 2.5~$\mathrm{M}$ nitrite microfluidic droplets as a function of droplet radius at P = 212 mW and a fit to $1/R^{2}$. (F) Linear fit to the combined data sets in (D) and (E) is used to determine the product of molar extinction coefficient ($\epsilon$) and photoionization quantum yield ($\phi$).}
\label{fig:1}
\end{figure*}

Micron-sized aqueous nitrite droplets are generated and stored in an atmospheric pressure quadrupole electrodynamic trap (QET) or microfluidic device (Fig. 1A). Both approaches, which differ only in the nature of the droplet interface, are described in previous publications,\cite{ref9,ref10,ref11,ref12,ref13} with additional experimental details in Sect. S1 of the Supplementary Information (SI). Once the droplets are formed, 532 nm light from a continuous wave (CW) laser is used both to excite the photochemistry in a single droplet and to simultaneously record Raman spectra as the reaction proceeds. As shown in Figs. 1B and 1C over the timescales of minutes to hours the peak at 1335 $\mathrm{cm}^{-1}$ corresponding to nitrite decreases, accompanied by the formation of a nitrate band at 1055 $\mathrm{cm}^{-1}$. The photoreaction is conducted one droplet at a time, so once the photochemistry in one droplet is completed, a second droplet is illuminated, starting the photochemical reaction again. This control ensures that photochemical reactions are driving the transformation rather than a dark reaction with a trace gas or liquid phase contaminant in the QET or microfluidic chip. To check for the presence of photoactive impurities such as trace metals, the photochemical kinetics are measured using both reagent grade (Sigma Aldrich, 901903, Reagent Plus, $\geq$ 90\%) and high purity (Sigma Aldrich 563218, 99.999\% trace metal basis) nitrite. The kinetics observed for both grades are identical, indicating that trace metal contamination is not responsible for the observed photochemistry. In the QET where the gas phase composition is easily changed from pure \ce{N2} to air, a $\sim$50-70\% increase in photochemical formation rate of nitrate is observed indicating that although dissolved \ce{O2} might play a role in the downstream chemistry, it is not an essential component for the photochemical reaction to occur. There is no evidence that the reaction rate depends upon the excess charge on droplets trapped in the QET.

The observed photochemical decay rate ($k_{\mathrm{obs}}$) in microfluidic droplets is obtained by fitting an exponential function to the decay and formation kinetics of nitrite and nitrate. $k_{\mathrm{obs}}$ increases linearly with laser power (P) for two different droplet sizes (Fig. 1D), with a slope that is larger for smaller droplet sizes. Linear scaling indicates a photochemical mechanism not driven by two photon or other nonlinear processes. Shown in Fig. 1E is $k_{\mathrm{obs}}$ for a range of droplet sizes at fixed P = 212 mW. The rate scales as $R^{-2}$ and is consistent with the scaling derived by Logozzo and Preston\cite{ref14} for a photochemical reaction that occurs within the droplet's volume, and not consistent with surface photochemistry, which scales as $R^{-3}$.

As derived in Sect. S2 of the SI, the photon flux into the droplet scales as $R^{-2}$ when a small Gaussian laser beam illuminates a larger droplet. If photochemical initiation is rate limiting, then $k_{\mathrm{obs}}$ depends only on the photon flux, molar extinction coefficient ($\epsilon$) and quantum yield ($\phi)$,

\begin{equation}
k_{\mathrm{obs}}=\frac{P}{R^{2}}\left[\frac{3\lambda}{2\pi h c}
\left(\frac{1000\ln(10)}{N_{\mathrm{A}}}\right)\right]\epsilon\phi ,
\label{eq:kobs}
\end{equation}

where $\lambda$ is the laser wavelength, c the speed of light, h Planck's constant and $N_{\mathrm{A}}$ is Avogadro's number. Fig. 1F collapses the data from the varying laser powers (Fig.~1D) and droplet sizes (Fig. 1E) onto the predicted linear relationship of Eq.~\eqref{eq:kobs} with a slope equal to $\epsilon \phi$ = $(3.46\pm0.19)\times10^{-6}$ $\mathrm{M}^{-1}\,\mathrm{cm}^{-1}$. For a quantum yield of 1, this would correspond to an absorption cross section ($\sigma$) of $1.3\times10^{-26}$ $\mathrm{cm}^{2}\,\mathrm{molecule}^{-1}$, which is 7 orders of magnitude smaller than the peak absorption of nitrite at 360 nm and well below the detection limit of conventional UV-Vis spectrometers. This exquisite sensitivity to very low probability events in microdroplets is a consequence of being able to focus a moderately powered CW laser with beam waist of 0.59 $\mu$m onto a small reaction volume. For example, focusing a 100 mW laser into a 10 $\mu$m droplet generates a flux into the sample volume of $8\times10^{3}$ $\mathrm{W}\,\mathrm{cm}^{-2}$, whereas for a cuvette-sized container, the flux into the sample volume drops precipitously to $\sim$$10^{-2}$ $\mathrm{W}\,\mathrm{cm}^{-2}$.

\begin{figure*}[t]
\centering
\includegraphics[width=\textwidth]{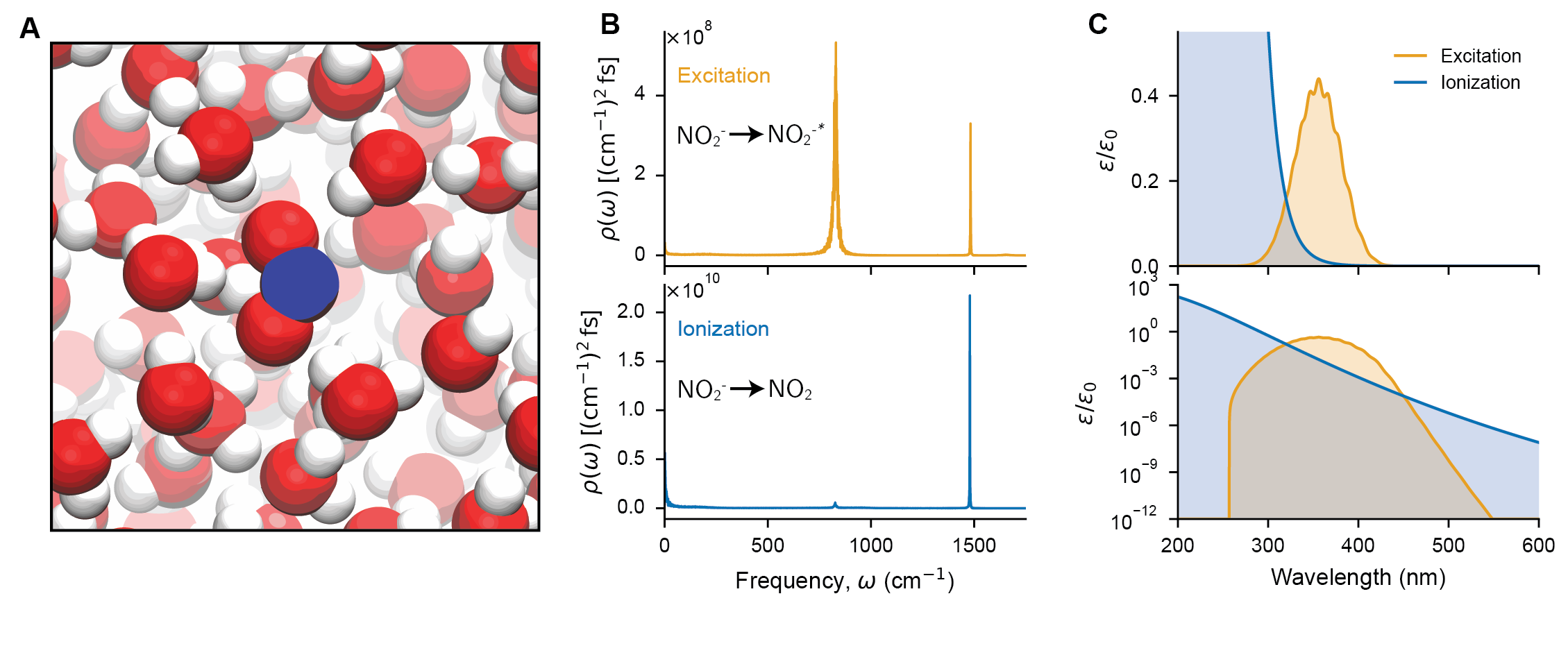}
\caption{Molecular picture and computed spectroscopic signatures of nitrite excitation and ionization in aqueous solution. (A) Representative molecular-dynamics snapshot of a nitrite ion hydrogen-bonded to surrounding water molecules (nitrogen: blue, oxygen: red, hydrogen: white). The first solvation shell is highlighted, while more distant solvent is shown with reduced opacity. (B) Spectral distributions $\rho(\omega)$ derived from energy-gap fluctuations for electronic excitation (top) and ionization (bottom), illustrating the contributions of solvent modes at different frequencies $\omega$ to the two processes. The overall scale is proportional to the reorganization energy and indicates the coupling strength to the solvent. (C) Corresponding calculated optical spectra for excitation and ionization, shown on linear (top) and logarithmic (bottom) intensity scales to emphasize the much broader long-wavelength tail of the ionization process. Each spectrum is individually referenced to an experimentally derived intensity that includes effects of quantum yield and intrinsic absorption $\epsilon_{0}$.}
\label{fig:2}
\end{figure*}
While the experimental data clearly implicate a photochemical process far detuned from known transitions, molecular dynamics simulations are employed to assess the relative feasibility of excitation to the $n\rightarrow\pi^{*}$ band and direct photoionization under 532-nm illumination. Excitation into the band centered near 360 nm initiates established pathways that generate $\mathrm{NO}$ and $\mathrm{OH}^{\bullet}$,\cite{ref1} whereas photoionization creates $\mathrm{NO}_{2}$ and a hydrated electron, ($e_{\mathrm{aq}}^{-}$), and therefore initiates distinct solution-phase chemistry.\cite{ref15} Although solvation blue-shifts the mean ionization threshold relative to the gas phase,\cite{ref16,ref17} rare solvent fluctuations could transiently reduce the ionization energy gap and enable absorption at substantially longer wavelengths. To quantify these spectral tails, we constructed bespoke, state-specific force fields for ground-state $\mathrm{NO}_{2}^{-}$, excited $\mathrm{NO}_{2}^{-*}$, and $\mathrm{NO}_{2}$ in an aqueous solution\cite{ref18} and sampled the corresponding vertical energy gaps along equilibrium ground-state trajectories with explicit solvation (Fig. 2A). The magnitude and time scales of the solvent fluctuations are encoded in the energy-gap autocorrelation functions, or equivalently their Fourier transform spectral distributions $\rho(\omega)$, which provide mode-specific information on the coupling of the electronic excitation to environmental modes. From $\rho\left(\omega\right)$, absorption spectra are calculated assuming the optical response is Gaussian, which is expected to be valid for vibrational models and collective solvent polarization.\cite{ref19} Details of the force fields, spectral reconstruction, and calibration of the energy gap are provided in Sect. S3 of the SI and in Figs. S2-S7.

The simulations reveal markedly different solvent responses for the two photochemical pathways (Fig. 2B). Photoexcitation to the $n\rightarrow\pi^{*}$ state produces only a modest charge redistribution coupling predominantly to intramolecular modes of nitrite at $\sim$1000 $\mathrm{cm}^{-1}$, yielding a narrow band centered near 360 nm consistent with previously reported UV absorption spectra of nitrite.\cite{ref1} In contrast, photoionization transfers the excess electron from nitrite to the surrounding water and requires a much larger collective solvent reorganization, which couples far more strongly to low-frequency solvent motions, broadening the spectrum and generating a much more pronounced long-wavelength tail (Fig. 2C). This tail is phenomenologically analogous to the Urbach tail observed in polar semiconductors\cite{ref20} that extends the conduction band edge into the optical band gap. At 532 nm, the long absorption tail of photoionization produces a cross section that is orders of magnitude larger than that of the $n\rightarrow\pi^{*}$ photoexcitation. Although the absolute magnitude of the far-red tail depends on how the amplitudes are normalized, the much larger absorption cross section for photoionization relative to photoexcitation at 532 nm is general across a range of choices. Using known experimental features in the optical absorption of nitrite (see Sect. S3) to calibrate the theoretical model, yields a product of the extinction coefficient and its corresponding quantum yield at 532 nm of $\epsilon$ $\phi$ = $1.38\times10^{-6}$ $\mathrm{M}^{-1}\,\mathrm{cm}^{-1}$, thus providing strong evidence that the experimental measurements are probing photoionization rather than photoexcitation pathways of aqueous nitrite. These calculations strongly implicate photoionization as the likely photochemical pathway observed in the microdroplets.

\begin{figure*}[t]
\centering
\includegraphics[width=\textwidth]{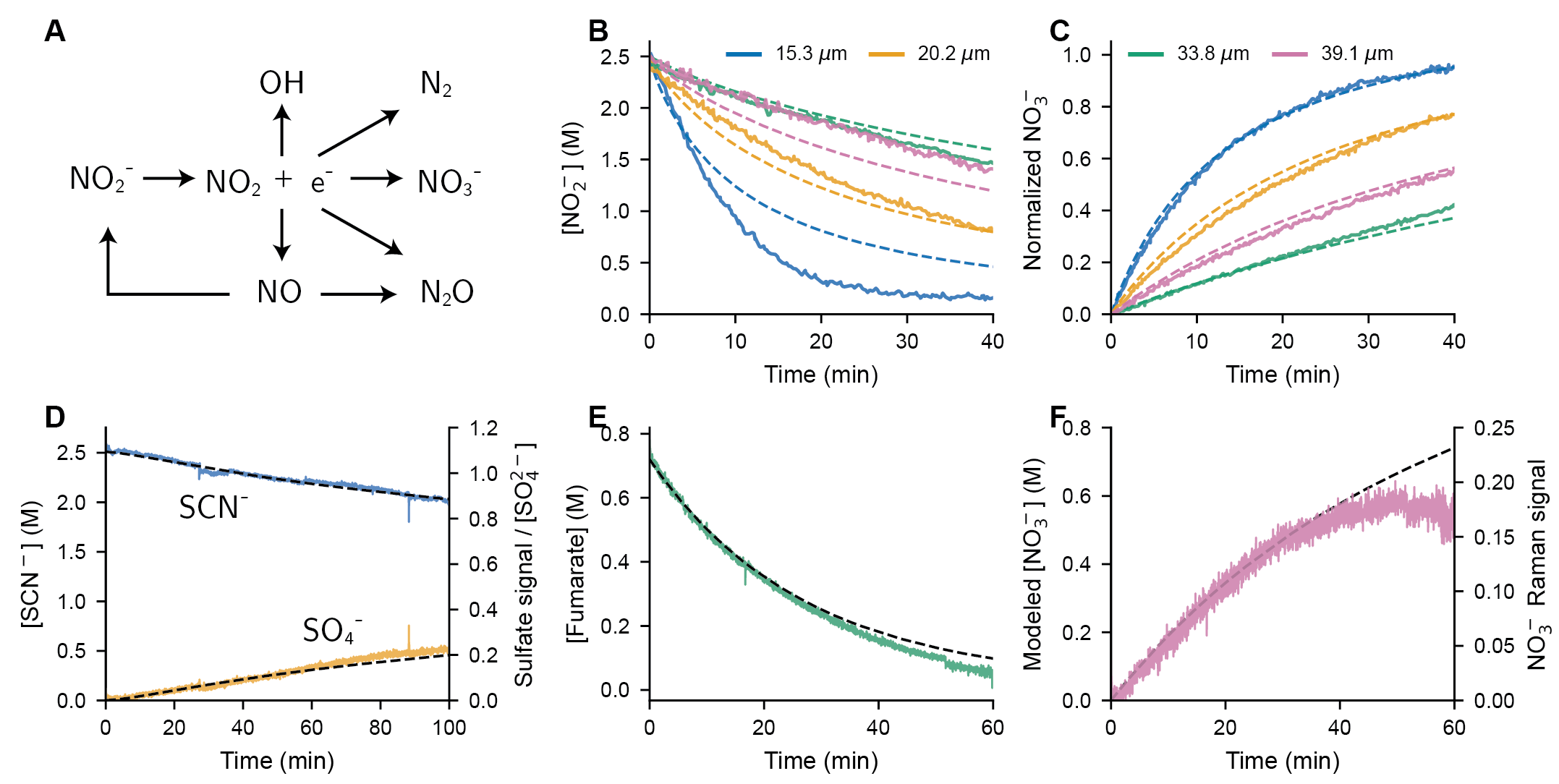}
\caption{Kinetic modeling and radical scavenger measurements are consistent with a solvated electron intermediate formed from photoionization of nitrite solutions. (A) Skeletal photoionization mechanism showing pathways that connect major products and intermediates. Experimental (solid lines) and simulated (dashed lines) (B) nitrite decay kinetics and (C) nitrate formation kinetics. Simulated and experimental nitrate formation kinetics are compared by normalizing each at t = 3000 s. (D) Measured (solid lines) and simulated (dashed lines) \ce{SCN^-} scavenger kinetics for a droplet size of R = 32 $\mu$m and initial composition of [\ce{SCN^-}] = 2.51~$\mathrm{M}$ and [$\mathrm{NO}_{2}^{-}$] = 2.5~$\mathrm{M}$. Simulated sulfate concentration (left axis) and measured sulfate Raman band intensity (right axis) vs. reaction time. (E) Measured (solid lines) and simulated (dashed lines) scavenger kinetics for R = 33 $\mu$m droplets with initial concentrations of [Fumarate] = 0.72~$\mathrm{M}$ and [$\mathrm{NO}_{2}^{-}$] = 2.38~$\mathrm{M}$ at pH = 6. (F) Simulated formation kinetics of nitrate (left axis, dashed line) vs. nitrate Raman signal (right axis, solid line) vs. reaction time. The kinetics shown in (B)--(F) are an average of 3 replicates.}
\label{fig:3}
\end{figure*}
To confirm that photoionization is indeed a viable mechanism for nitrate production in microdroplets, we constructed a kinetic model of aqueous phase photoionization. Most of the elementary steps and rate constants in the model have been previously examined and quantified in pulsed radiolysis studies, and the sole unknown for application to our system is the rate of photoionization.\cite{ref15} The mechanism (shown explicitly in Table S3) has four main elements: photochemistry, intermediate pathways, nitrate formation and scavenger reactions, as discussed below and summarized in the simplified reaction network shown in Fig. 3A. Photoionization of $\mathrm{NO}_{2}^{-}$ produces $\mathrm{NO}_{2}$ and $e_{\mathrm{aq}}^{-}$ as intermediates before subsequent conversion to nitrate or regeneration of $\mathrm{NO}_{2}^{-}$. The rate of this step is computed using Eq.~\eqref{eq:kobs}, for a given droplet size and laser power. For comparison, Fig. S8 shows the reaction pathways produced by photoexcitation to the $n\rightarrow\pi^{*}$ state, which generates \ce{NO} and $\mathrm{OH}^{\bullet}$ but no solvated electron.\cite{ref1}

The solvated electron reacts rapidly with nitrite and nitrate, which constitute the major sinks under our experimental conditions, while $e_{\mathrm{aq}}^{-}$ also reacts with other products and intermediates such as \ce{N2O}, \ce{NO}, and \ce{NO2}. Most of these reactions are fast, proceeding at or near the diffusion limit. Shown in Fig. 3B and 3C are the kinetic model predictions compared to the experimental photochemical kinetics recorded as a function of microfluidic droplet radius. Kinetics recorded for a macroscopic sample obtained by filling a 40000 x 800 x 80 $\mu$m channel in a microfluidic chip (e.g. Cellix Vena 8) with an aqueous nitrite solution showed that the solution is stable under illumination for many hours (beyond the timescale shown in Fig. 3B) and we observe no spectral evidence that nitrate is formed indicating that there is a substantial acceleration of the photochemical kinetics in microdroplets. Shown in Fig. 3C are the formation kinetics of nitrate, which mirror the loss of nitrite. The kinetic simulations reasonably capture the size dependent decay kinetics of nitrite as well as the formation kinetics of nitrate. The experimental decay kinetics of nitrite are approximately exponential, whereas the simulated kinetics are non-exponential with nitrite consumption slowing at longer times. The rate of nitrite decay at long times is quite sensitive to the $e_{\mathrm{aq}}^{-}$ sinks in the model. For example, the aqueous droplets are surrounded by fluorinated oil and surfactant, both of which are reactive towards $e_{\mathrm{aq}}^{-}$ but not included in the model. Additionally, we have not included any steps to account for mass transfer\cite{ref11} of molecules out of the aqueous phase into the oil, which is a likely loss mechanism especially for species such as \ce{NO}, \ce{N2O}, or \ce{NO2}. Nevertheless, given these model limitations, the kinetic model replicates the reaction time scales and droplet size dependence for nitrite and nitrate observed experimentally.

To probe the existence of the solvated electron intermediate, additional experiments are conducted with $\mathrm{OH}^{\bullet}$ and $e_{\mathrm{aq}}^{-}$ scavengers, fumarate (\ce{C4H4O4}) and thiocyanate (\ce{SCN^-}). These scavengers, used in separate droplet populations, exhibit differential reactivity towards $\mathrm{OH}^{\bullet}$ and $e_{\mathrm{aq}}^{-}$. Fumarate reacts rapidly with both $\mathrm{OH}^{\bullet}$ and $e_{\mathrm{aq}}^{-}$ with rate coefficients of $5\times10^{9}$ $\mathrm{M}^{-1}\,\mathrm{s}^{-1}$ and $2.2\times10^{10}$ $\mathrm{M}^{-1}\,\mathrm{s}^{-1}$, respectively,\cite{ref21} whereas, \ce{SCN^-} is highly reactive towards $\mathrm{OH}^{\bullet}$ ($1.4\times10^{10}$ $\mathrm{M}^{-1}\,\mathrm{s}^{-1}$)\cite{ref15,ref22} but not $e_{\mathrm{aq}}^{-}$ ($<10^{6}$ $\mathrm{M}^{-1}\,\mathrm{s}^{-1}$).\cite{ref15} Shown in Fig. 3 are the decay kinetics of \ce{SCN^-} and fumarate extracted from time dependent Raman spectra (Fig. S9) and described in Sect. S5 of the SI. In Fig. 3D the decay kinetics of \ce{SCN^-} and the associated product formation kinetics of sulfate, a terminal product of \ce{SCN^-} oxidation, are replicated, indicating that the $\mathrm{OH}^{\bullet}$ sources and sinks in our kinetic model are reasonably accurate. The decay of fumarate is also well replicated by the kinetic model (Figs. 3E and 3F) as well as the accompanying formation kinetics of nitrate. Turning off the reaction of fumarate with $e_{\mathrm{aq}}^{-}$ produces model predictions that are much too slow compared to experiment indicating that the main sink for fumarate in the droplet is its reaction with solvated electrons---a distinguishing feature of photoionization.

The kinetic model of photoionization explains previously measured nitrite reaction kinetics in laser trapped microdroplets. In aerosol optical tweezers, Meroni \textit{et al.}\cite{ref23} report nitrate formation kinetics whereas Monroe \textit{et al.}\cite{ref24} report nitrite decay kinetics. Both the formation kinetics of nitrate and the nitrite decay rate are replicated by our photoionization mechanism (Fig. S10), based on estimates of the laser power and droplet size in these prior experiments. Although our photoionization mechanism provides a reasonable and quantitative account of these previous observations, the authors of these prior studies offer a rather different interpretation of their data as discussed in Sect. 5 of the SI.

While the cross section is low, the broad absorption band for photoionization of nitrite is predicted to have strong overlap with the visible solar spectrum. To assess the potential environmental contribution of photoexcitation and photoionization under broadband solar illumination, we calculated pathway-specific effective photochemical rates, $J=\int \sigma\left(\lambda\right) F\left(\lambda\right) \phi\left(\lambda\right) d\lambda$, where $\sigma\left(\lambda\right)$ is the simulated cross section, $F\left(\lambda\right)$ is the spectral photon irradiance derived from the ASTM G-173 AM1.5 global-tilted terrestrial reference spectrum.\cite{ref25} $\phi\left(\lambda\right)$ is the corresponding product quantum yield of $e_{\mathrm{aq}}^{-}$ for photoionization (PI) and $\mathrm{OH}^{\bullet}$ for photoexcitation (PE). Using the computed form $\sigma\left(\lambda\right)$ from the molecular simulations, calibrated with the experimentally determined value of $\epsilon \phi$ at 532 nm, yields $J_{PI}=2.2 \times 10^{-6}$\,$\mathrm{s}^{-1}$. For photoexcitation, $J_{PE}=1.0 \times 10^{-5}$\,$\mathrm{s}^{-1}$, assuming a wavelength independent quantum yield for $\mathrm{OH}^{\bullet}$ production of 2\%.\cite{ref26,ref27,ref28,ref29} The photoexcitation rate computed here is reasonably consistent with previous reports\cite{ref28} under matched assumptions for the $\mathrm{OH}^{\bullet}$ quantum yield and droplet enhancement factor. Droplet enhancement factors include light-scattering and multiple internal reflections of the actinic flux that increase photochemical rates in droplets by $1.6$--$2.6\times$.\cite{ref30,ref31}

These estimates suggest that photoionization of aqueous nitrite could represent a previously unrecognized source of nitrogen oxides and solvated electrons in atmospheric aerosols and natural waters. At a representative environmental nitrite concentration of $3\times10^{-7}$~$\mathrm{M}$ , photoionization produces \ce{NO2} and solvated electrons at $7\times10^{-13}$ $\mathrm{M}\,\mathrm{s}^{-1}$, approximately 22\% of the $\mathrm{OH}^{\bullet}$/\ce{NO} production rate from photoexcitation of $3\times10^{-12}$ $\mathrm{M}\,\mathrm{s}^{-1}$. The subsequent fate of the solvated electron depends on solution composition, but under dilute aerobic conditions its dominant sink is expected to be dissolved \ce{O2}, producing superoxide and hydroperoxyl radicals that can drive additional oxidation and H-atom abstraction chemistry. Solvated electrons may also form H atoms under acidic conditions or directly reduce unsaturated organic molecules.

These primary photochemical rates do not, however, determine the environmental lifetime of nitrite because downstream reactions can regenerate nitrite. We therefore simulated both mechanisms including reaction pathways relevant to dilute aqueous environments (Tables S3--S5). At 260 nM nitrite, photoionization consumes nitrite approximately three times faster than photoexcitation, whereas at millimolar concentrations photoexcitation is approximately five times faster (Figs. S11A and B). This reversal arises from concentration-dependent regeneration pathways. Under dilute conditions, \ce{O2} efficiently scavenges solvated electrons, suppressing reactions such as electron--\ce{NO2} recombination that regenerate nitrite. Photoexcitation likewise contains regeneration pathways whose importance depends on the fate of $\mathrm{OH}^{\bullet}$ radicals. Thus, despite its smaller primary photochemical rate, photoionization can become the dominant abiotic sink for nitrite under environmentally relevant dilute conditions.

More generally, these results point to a new class of photochemical transformation that may drive important atmospheric and aquatic environmental chemistry. Although photoionization at visible wavelengths is expected to have very small absolute cross sections, the long spectral tail arising from solvent coupling produces substantial overlap with the terrestrial solar spectrum enabling these rare events to have unexpected impacts on photochemical lifetimes. Indeed, recent studies implicate photoionization to rationalize anomalous photochemistry in concentrated sulfate solutions, though without clear mechanistic explanation.\cite{ref32,ref33} Future work will need to focus on refining the accuracy of the theoretical models used here, which when combined with the photochemical sensitivity achieved using microdroplets, will enable a more comprehensive survey of how important photoionization pathways are driving chemistry in atmospheric and environmental systems.

\begin{acknowledgments}
This experimental work was supported by the Condensed Phase and Interfacial Molecular Science Program (CPIMS), in the Chemical Sciences Geosciences and Biosciences Division of the Office of Basic Energy Sciences of the U.S. Department of Energy under Contract No. DE-AC02-05CH11231. The theory was supported by Laboratory Directed Research and Development Program of Lawrence Berkeley National Laboratory under U.S. Department of Energy Contract No. DE-AC02-05CH11231. AA and BJW were partially supported by the William M. Keck Foundation (8959).
\end{acknowledgments}


\begin{thebibliography}{0}%
\makeatletter
\providecommand \@ifxundefined [1]{%
 \@ifx{#1\undefined}
}%
\providecommand \@ifnum [1]{%
 \ifnum #1\expandafter \@firstoftwo
 \else \expandafter \@secondoftwo
 \fi
}%
\providecommand \@ifx [1]{%
 \ifx #1\expandafter \@firstoftwo
 \else \expandafter \@secondoftwo
 \fi
}%
\providecommand \natexlab [1]{#1}%
\providecommand \enquote  [1]{``#1''}%
\providecommand \bibnamefont  [1]{#1}%
\providecommand \bibfnamefont [1]{#1}%
\providecommand \citenamefont [1]{#1}%
\providecommand \href@noop [0]{\@secondoftwo}%
\providecommand \href [0]{\begingroup \@sanitize@url \@href}%
\providecommand \@href[1]{\@@startlink{#1}\@@href}%
\providecommand \@@href[1]{\endgroup#1\@@endlink}%
\providecommand \@sanitize@url [0]{\catcode `\\12\catcode `\$12\catcode
  `\&12\catcode `\#12\catcode `\^12\catcode `\_12\catcode `\%12\relax}%
\providecommand \@@startlink[1]{}%
\providecommand \@@endlink[0]{}%
\providecommand \url  [0]{\begingroup\@sanitize@url \@url }%
\providecommand \@url [1]{\endgroup\@href {#1}{\urlprefix }}%
\providecommand \urlprefix  [0]{URL }%
\providecommand \Eprint [0]{\href }%
\providecommand \doibase [0]{https://doi.org/}%
\providecommand \selectlanguage [0]{\@gobble}%
\providecommand \bibinfo  [0]{\@secondoftwo}%
\providecommand \bibfield  [0]{\@secondoftwo}%
\providecommand \translation [1]{[#1]}%
\providecommand \BibitemOpen [0]{}%
\providecommand \bibitemStop [0]{}%
\providecommand \bibitemNoStop [0]{.\EOS\space}%
\providecommand \EOS [0]{\spacefactor3000\relax}%
\providecommand \BibitemShut  [1]{\csname bibitem#1\endcsname}%
\let\auto@bib@innerbib\@empty
%</preamble>
\end{thebibliography}%




\begin{thebibliography}{33}

\bibitem{ref1}
J. Mack, J. R. Bolton, Photochemistry of nitrite and nitrate in aqueous solution: a review. Journal of Photochemistry and Photobiology A: Chemistry 128, 1-13 (1999).

\bibitem{ref2}
T. A. Doane, The Abiotic Nitrogen Cycle. ACS Earth and Space Chemistry 1, 411-421 (2017).

\bibitem{ref3}
D. E. Canfield, A. N. Glazer, P. G. Falkowski, The Evolution and Future of Earth's Nitrogen Cycle. Science 330, 192-196 (2010).

\bibitem{ref4}
S. Gligorovski, R. Strekowski, S. Barbati, D. Vione, Environmental Implications of Hydroxyl Radicals ($\bullet$OH). Chemical Reviews 115, 13051-13092 (2015).

\bibitem{ref5}
C. George, M. Ammann, B. D'Anna, D. J. Donaldson, S. A. Nizkorodov, Heterogeneous Photochemistry in the Atmosphere. Chemical Reviews 115, 4218-4258 (2015).

\bibitem{ref6}
K. Mopper, X. Zhou, Hydroxyl Radical Photoproduction in the Sea and Its Potential Impact on Marine Processes. Science 250, 661-664 (1990).

\bibitem{ref7}
O. C. Zafiriou, M. B. True, Nitrite photolysis in seawater by sunlight. Marine Chemistry 8, 9-32 (1979).

\bibitem{ref8}
D. Vione, V. Maurino, C. Minero, E. Pelizzetti, in Environmental Photochemistry Part II, P. Boule, D. W. Bahnemann, P. K. J. Robertson, Eds. (Springer Berlin Heidelberg, Berlin, Heidelberg, 2005), pp. 221-253.

\bibitem{ref9}
M. I. Jacobs \textit{et al.}, Exploring Chemistry in Microcompartments Using Guided Droplet Collisions in a Branched Quadrupole Trap Coupled to a Single Droplet, Paper Spray Mass Spectrometer. Analytical Chemistry 89, 12511-12519 (2017).

\bibitem{ref10}
A. M. Prophet, K. Polley, G. J. Van Berkel, D. T. Limmer, K. R. Wilson, Iodide oxidation by ozone at the surface of aqueous microdroplets. Chemical Science 15, 736-756 (2024).

\bibitem{ref11}
H. Row, B. J. Wallace, J. B. Fernandes, K. R. Wilson, K. K. Mandadapu, Interfacial mass transfer resistance at fluid-fluid interfaces. arXiv preprint arXiv:2606.14678, (2026).

\bibitem{ref12}
B. J. Wallace \textit{et al.}, Enhanced reactivity at the oil--water interface accelerates the synthesis of zymonic acid in microemulsions. Chemical Science 16, 15155-15165 (2025).

\bibitem{ref13}
M. D. Willis, G. Rovelli, K. R. Wilson, Combining Mass Spectrometry of Picoliter Samples with a Multicompartment Electrodynamic Trap for Probing the Chemistry of Droplet Arrays. Analytical Chemistry 92, 11943-11952 (2020).

\bibitem{ref14}
A. Logozzo, T. C. Preston, Photochemical transformation of weakly absorbing organics by visible light in microdroplets. Physical Chemistry Chemical Physics 28, 1207-1221 (2026).

\bibitem{ref15}
G. V. Buxton, C. L. Greenstock, W. P. Helman, A. B. Ross, Critical review of rate constants for reactions of hydrated electrons, hydrogen atoms and hydroxyl radicals ($\bullet\mathrm{OH}/\bullet\mathrm{O}^{-}$) in aqueous solution. Journal of Physical and Chemical Reference Data 17, 513-886 (1988).

\bibitem{ref16}
E. Herbst, T. A. Patterson, W. C. Lineberger, Laser photodetachment of $\mathrm{NO}_{2}^{-}$. The Journal of Chemical Physics 61, 1300-1304 (1974).

\bibitem{ref17}
J. H. Richardson, L. M. Stephenson, J. I. Brauman, Photodetachment of $\mathrm{NO}_{2}^{-}$; experimental evidence for a new isomer. Chemical Physics Letters 25, 318-320 (1974).

\bibitem{ref18}
T. P. Fay, D. T. Limmer, Unraveling the mechanisms of triplet state formation in a heavy-atom free photosensitizer. Chemical Science 15, 6726-6737 (2024).

\bibitem{ref19}
T. J. Zuehlsdorff, A. Montoya-Castillo, J. A. Napoli, T. E. Markland, C. M. Isborn, Optical spectra in the condensed phase: Capturing anharmonic and vibronic features using dynamic and static approaches. The Journal of Chemical Physics 151, (2019).

\bibitem{ref20}
J. D. Dow, D. Redfield, Toward a Unified Theory of Urbach's Rule and Exponential Absorption Edges. Physical Review B 5, 594-610 (1972).

\bibitem{ref21}
Á. Bíró, L. Wojnárovits, Reaction of diethyl maleate and diethyl fumarate with hydrated electrons and hydroxyl radicals. Radiation Physics and Chemistry 47, 389-392 (1996).

\bibitem{ref22}
B. H. Milosavljevic, J. A. LaVerne, Pulse Radiolysis of Aqueous Thiocyanate Solution. The Journal of Physical Chemistry A 109, 165-168 (2005).

\bibitem{ref23}
C. Meroni, A. Gandolfo, C. George, Nitrite to Nitrate Spontaneous Oxidation in Suspended Aqueous Droplets. Geophysical Research Letters 53, e2025GL120957 (2026).

\bibitem{ref24}
L. W. Monroe, J. W. Hall, G. M. Thornhill, R. C. Sullivan, Spontaneous Oxidation of Nitrous Acid to Nitric Acid in Supermicron Aqueous Droplets Is Acid-Accelerated. ACS Earth and Space Chemistry 9, 1152-1164 (2025).

\bibitem{ref25}
ASTM International, ASTM G173-03(2020): Standard Tables for Reference Solar Spectral Irradiances: Direct Normal and Hemispherical on 37$^{\circ}$ Tilted Surface (ASTM International, West Conshohocken, PA, 2020). DOI: \href{https://doi.org/10.1520/G0173-03R20}{10.1520/G0173-03R20}

\bibitem{ref26}
A. Alif, P. Boule, Photochemistry and environment Part XIV. Phototransformation of nitrophenols induced by excitation of nitrite and nitrate ions. Journal of Photochemistry and Photobiology A: Chemistry 59, 357-367 (1991).

\bibitem{ref27}
O. C. Zafiriou, R. Bonneau, Wavelength-dependent quantum yield of $\mathrm{OH}^{\bullet}$ radical formation from photolysis of nitrite ion in water. Photochemistry and Photobiology 45, 723-727 (1987).

\bibitem{ref28}
R. Zellner, M. Exner, H. Herrmann, Absolute $\mathrm{OH}^{\bullet}$ quantum yields in the laser photolysis of nitrate, nitrite and dissolved \ce{H2O2} at 308 and 351 nm in the temperature range 278--353 K. Journal of Atmospheric Chemistry 10, 411-425 (1990).

\bibitem{ref29}
H. Herrmann, On the photolysis of simple anions and neutral molecules as sources of $\mathrm{O}^{-}/\mathrm{OH}$, $\mathrm{SO}_{x}^{-}$, and Cl in aqueous solution. Physical Chemistry Chemical Physics 9, 3935-3964 (2007).

\bibitem{ref30}
S. Madronich, Photodissociation in the atmosphere: 1. Actinic flux and the effects of ground reflections and clouds. Journal of Geophysical Research: Atmospheres 92, 9740-9752 (1987).

\bibitem{ref31}
P. Nissenson, C. J. H. Knox, B. J. Finlayson-Pitts, L. F. Phillips, D. Dabdub, Enhanced photolysis in aerosols: evidence for important surface effects. Physical Chemistry Chemical Physics 8, 4700-4710 (2006).

\bibitem{ref32}
J. D. Cope, K. H. Bates, L. N. Tran, K. A. Abellar, T. B. Nguyen, Sulfur radical formation from the tropospheric irradiation of aqueous sulfate aerosols. Proceedings of the National Academy of Sciences 119, e2202857119 (2022).

\bibitem{ref33}
A. Petersen, Z. Luo, A. Wong, R. Xiao, T. B. Nguyen, Direct Photolysis and Photosensitized Reactions of Sulfate Ions in Aerosol Water Form Radical Oxidants, Secondary Brown Carbon, and Organosulfates. Environmental Science \& Technology, (2026).

\end{thebibliography}
\end{document}

\title{\texorpdfstring{Supplementary Information for\\Solvent Fluctuations Enable Photoionization of Aqueous Nitrite Microdroplets}{Supplementary Information for Solvent Fluctuations Enable Photoionization of Aqueous Nitrite Microdroplets}}

\author{Brandon J. Wallace}
\affiliation{Chemical Sciences Division, Lawrence Berkeley National Laboratory, Berkeley, CA 94720, USA}
\affiliation{Department of Chemistry, University of California, Berkeley, CA 94720, USA}

\author{Eric R. Heller}
\affiliation{Chemical Sciences Division, Lawrence Berkeley National Laboratory, Berkeley, CA 94720, USA}
\affiliation{Department of Chemistry, University of California, Berkeley, CA 94720, USA}
\affiliation{Materials Sciences Division, Lawrence Berkeley National Laboratory, Berkeley, CA 94720, USA}

\author{Alex Prophet}
\affiliation{Chemical Sciences Division, Lawrence Berkeley National Laboratory, Berkeley, CA 94720, USA}

\author{Pranav Ballapuram}
\affiliation{Chemical Sciences Division, Lawrence Berkeley National Laboratory, Berkeley, CA 94720, USA}
\affiliation{Department of Chemistry, University of California, Berkeley, CA 94720, USA}

\author{Ashok Ajoy}
\affiliation{Chemical Sciences Division, Lawrence Berkeley National Laboratory, Berkeley, CA 94720, USA}
\affiliation{Department of Chemistry, University of California, Berkeley, CA 94720, USA}

\author{David T. Limmer}
\affiliation{Chemical Sciences Division, Lawrence Berkeley National Laboratory, Berkeley, CA 94720, USA}
\affiliation{Department of Chemistry, University of California, Berkeley, CA 94720, USA}
\affiliation{Materials Sciences Division, Lawrence Berkeley National Laboratory, Berkeley, CA 94720, USA}
\affiliation{Kavli Energy NanoScience Institute, Berkeley, CA 94720, USA}

\author{Kevin R. Wilson}
\affiliation{Chemical Sciences Division, Lawrence Berkeley National Laboratory, Berkeley, CA 94720, USA}

\maketitle

\section*{S1. Experimental Methods}

\emph{\textbf{A. Microfluidic Droplets:}} Micron-sized water-in-oil
emulsions are generated using commercial droplet generation chips
(Cellix, Droplet Generation Chip or ChipShop Fluidic 285). The chips are
pre-treated with a hydrophobic coating (Cellix, DropGen pre-coat). The
oil phase is HFE7500, a fluorinated oil, containing 2\% (by weight) of a
proprietary fluorosurfactant (RAN Biotechnologies). Emulsions are
generated using continuous flows of oil and aqueous solutions, which are
controlled using a microfluidic pump (CorSolutions PneuWave or Cellix
4U). The total flow rate for droplet generation is fixed between 8 - 10
\ensuremath{\mu}Lmin\textsuperscript{-1}. The ratio of the aqueous to oil flow rate
controls the final size of the droplets, where a smaller ratio produces
smaller emulsions. Once generated, the emulsions are stored in
commercial acrylic microfluidic chips with a glass coverslip (e.g.,
Vena8 with Glass Coverslips, Cellix), where the size and reproducibility
of emulsion droplets are assessed by imagining them with a commercial
inverted microscope (Olympus, IX70). To determine the size of the
emulsions from an image, pixels on the camera were calibrated using a
Stage Micrometer (Thorlabs). For example, emulsion droplets with an
average size of 26 \ensuremath{\mu}m, the standard deviation, in radius, is typically \ensuremath{\pm}
0.2 \ensuremath{\mu}m.

The photochemical kinetics are conducted using a commercial single-beam
optical trap (Biral, AOT-100). This instrument is originally designed to
trap single droplets in air, so several modifications are made to probe
emulsions as detailed in Wallace et al.\cite{ref1} The optical trap uses
a 532 nm laser, which is tightly focused onto the emulsion droplets
using a 100x objective. The beam waist at the focus, in the absence of a
microfluidic chip, is measured to be \textasciitilde{} 0.59 \ensuremath{\mu}m.
Backscattered Raman signals from the emulsions are collected and
directed into a spectrometer and CCD. A collimated LED light illuminates
the droplets for imaging and sizing purposes. Importantly, for these
studies, we measured the laser power at the sample location as function
of the power setting at the laser head. As shown in Fig. S1 there are
substantial losses from the internal optical elements of the Biral AOT
where the actual laser power interacting with the sample is roughly 25\%
of the output power setting on the laser. This is similar to the custom
AOT built by Sullivan and coworkers,\cite{ref2} where \textasciitilde70\%
of the laser light forming the optical trap is attenuated (i.e. lost) by
transfer optics.

\begin{figure}[htbp]
\centering
\includegraphics[width=0.58\linewidth]{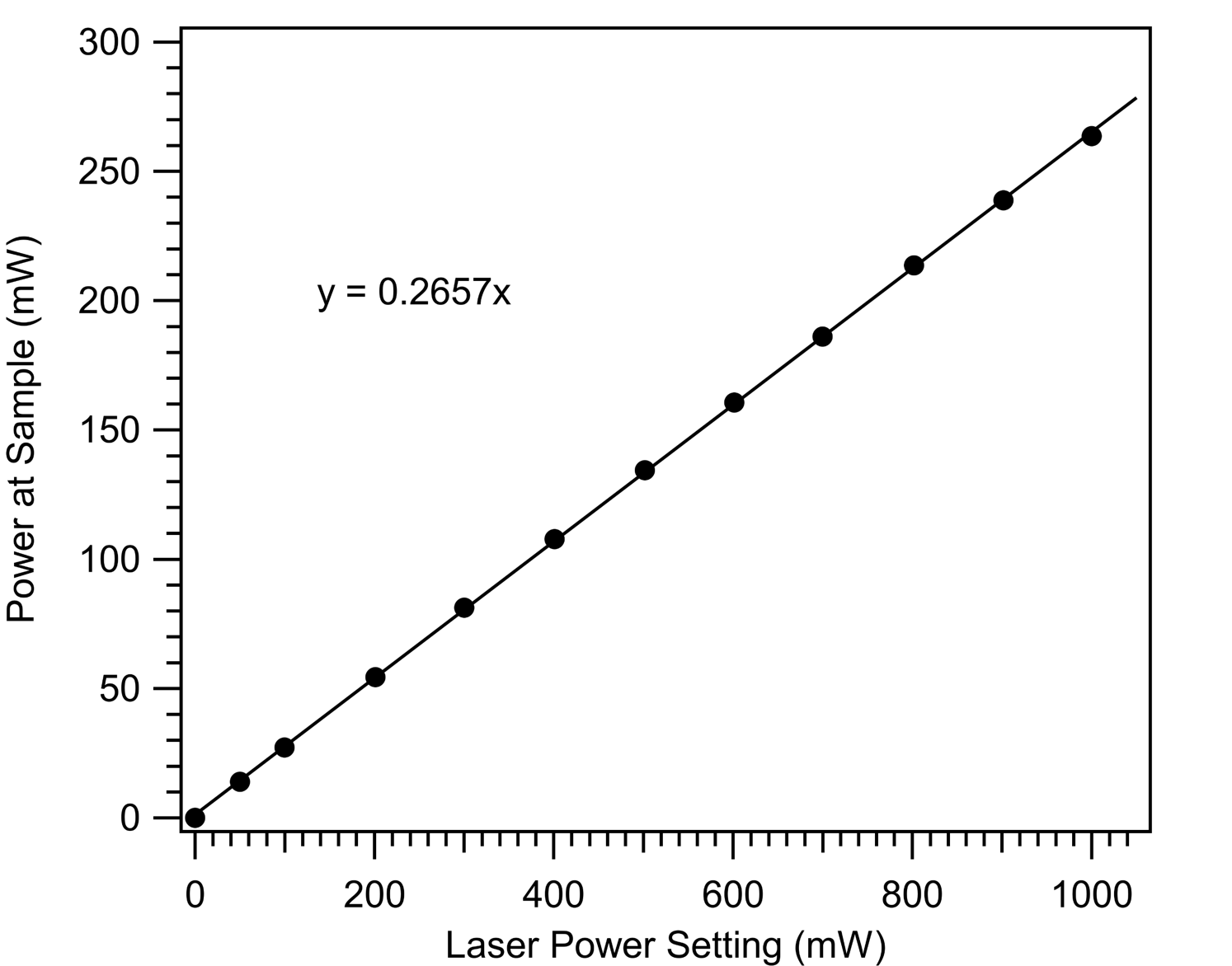}
\caption{Laser power at the sample vs. laser power setting for commercial Biral AOT-100 instrument (\url{https://www.biral.com/product/aot-100-aerosol-optical-tweezers/\#product-overview}).}
\label{fig:S1}
\end{figure}

\textbf{\emph{B. Quadrupole Electrodynamic Trap (QET):}} To verify the
broad applicability of the photochemistry observed in microemulsions,
for select experiments, we conducted photochemical reactions in nitrite
droplets dispensed and levitated in a QET. These parallel studies are
used to examine whether the nature of the droplet interface (air-water
vs. oil-water) influences the observed photochemistry. To do this we
used a smaller variant of a QET, described extensively in previous
publications.\cite{ref3,ref4,ref5} Light from a 532 nm CW laser (Coherent Verdi)
is focused on the levitated droplet using a long working distance
objective (10X Mitutoyo Plan Apochromatic). Raman scattering is
collected at 90\ensuremath{^{\circ}} to the incidence laser and fiber coupled to a CCD
spectrometer (Princeton Instruments IsoPlane).~ Although multiple
droplets are dispensed at time, experimental measurements are conducted
by illuminating only a single droplet and recording its Raman spectra as
a function of reaction time. As discussed in the main text, we do not
observe any large differences between the photochemical behavior of
microemulsions and microdroplets.

\section*{S2. Derivation of the Droplet-Averaged Photon Flux and Photochemical Rate}

We consider a Gaussian beam propagating along the \(z\)-axis through a
spherical droplet of radius \(R\). The beam is assumed to be centered on
the droplet. We would note that Logozzo and Preston\cite{ref6} report a
similar derivation as what follows.

\emph{\textbf{A. Local photon flux of a Gaussian beam:}} The local
intensity of a Gaussian beam is,

\begin{equation}
I(\rho,z) = \frac{2P}{\pi w(z)^{2}}\exp\left( - \frac{2\rho^{2}}{w(z)^{2}} \right),
\tag{S1}
\end{equation}
where \(P\) is the total laser power, \(\rho = \sqrt{x^{2} + y^{2}}\) is
the radial distance from the beam axis, and \(w(z)\) is the beam radius
at axial position \(z\). The local photon flux is,

\begin{equation}
F(\rho,z) = \frac{I(\rho,z)}{E_{ph}} = \frac{\lambda}{hc}I(\rho,z),
\tag{S2}
\end{equation}
where the photon energy is,

\begin{equation}
E_{ph} = \frac{hc}{\lambda}.
\tag{S3}
\end{equation}
Substituting the Gaussian intensity yields,

\begin{equation}
F(\rho,z) = \frac{2P\lambda}{\pi hc\, w(z)^{2}}\exp\left( - \frac{2\rho^{2}}{w(z)^{2}} \right).
\tag{S4}
\end{equation}
\emph{\textbf{B. Droplet-averaged photon flux:}} The droplet-averaged
photon flux is defined as,

\begin{equation}
\langle F\rangle = \frac{1}{V_{drop}}\int_{V_{drop}}^{}F(\mathbf{r})\, dV.
\tag{S5}
\end{equation}
For a sphere of radius \(R\),

\begin{equation}
V_{drop} = \frac{4}{3}\pi R^{3}.
\tag{S6}
\end{equation}
The beam is cylindrically symmetric about the \(z\)-axis, so the volume
element in cylindrical coordinates is,

\begin{equation}
dV = \rho\, d\rho\, d\theta\, dz.
\tag{S7}
\end{equation}
This volume element represents a cylinder slice of the beam of thickness
(\(dp\)) in the radial direction, angular width (\(d\theta\)), a
thickness \(dz\) along the beam and a circumference of \(\rho\).

At a given \(z\), the droplet cross-section has radius,

\begin{equation}
a(z) = \sqrt{R^{2} - z^{2}}.
\tag{S8}
\end{equation}
Thus, the volume integral is,

\begin{equation}
\int_{V_{drop}}^{}F(\mathbf{r})\, dV = \int_{- R}^{R}{\int_{0}^{2\pi}{\int_{0}^{\sqrt{R^{2} - z^{2}}}F}}(\rho,z)\,\rho\, d\rho\, d\theta\, dz.
\tag{S9}
\end{equation}
Physically, for every value of z, this integral is moving through all z
from one side of the droplet to the other (integral 1) and going all the
way around a circle (integral 2). Going from the beam center to the edge
of the droplet, where the gaussian beam may be clipped (integral 3).
Integrating over \(\theta\) gives,

\begin{equation}
\int_{V_{drop}}^{}F(\mathbf{r})\, dV = 2\pi\int_{- R}^{R}{\int_{0}^{\sqrt{R^{2} - z^{2}}}F}(\rho,z)\,\rho\, d\rho\, dz.
\tag{S10}
\end{equation}
This is because nothing depends on \(\theta\) because of symmetry, so it
simplifies to \(2\pi\). Substituting the Gaussian photon flux,

\begin{equation}
\int_{V_{drop}}^{}F(\mathbf{r})\, dV = 2\pi\int_{- R}^{R}{\int_{0}^{\sqrt{R^{2} - z^{2}}}\frac{2P\lambda}{\pi hc\, w(z)^{2}}}\exp\left( - \frac{2\rho^{2}}{w(z)^{2}} \right)\rho\, d\rho\, dz.
\tag{S11}
\end{equation}
and canceling \(\pi\) yields,

\begin{equation}
\int_{V_{drop}}^{}F(\mathbf{r})\, dV = \frac{4P\lambda}{hc}\int_{- R}^{R}\frac{1}{w(z)^{2}}\left\lbrack \int_{0}^{\sqrt{R^{2} - z^{2}}}\exp\left( - \frac{2\rho^{2}}{w(z)^{2}} \right)\rho\, d\rho \right\rbrack dz.
\tag{S12}
\end{equation}
At this stage we have separated the problem into a radial integral over
\(\rho\) and an axial integral over \(z\). In other words, we are first
deriving how much of the beam is captured at each cross-section, then
summing those slices along the droplet.

\emph{\textbf{C. Evaluation of the radial integral:}} To determine how
much of the beam is captured by the droplet at a particular z-slice.
Using,

\begin{equation}
\int\rho\exp\left( - \frac{2\rho^{2}}{w^{2}} \right)\, d\rho = - \frac{w^{2}}{4}\exp\left( - \frac{2\rho^{2}}{w^{2}} \right),
\tag{S13}
\end{equation}
we obtain,

\begin{equation}
\int_{0}^{a(z)}\exp\left( - \frac{2\rho^{2}}{w(z)^{2}} \right)\rho\, d\rho = \frac{w(z)^{2}}{4}\left\lbrack 1 - \exp\left( - \frac{2a(z)^{2}}{w(z)^{2}} \right) \right\rbrack.
\tag{S14}
\end{equation}
Substituting this back into the volume integral,

\begin{equation}
\begin{aligned}
\int_{V_{drop}}^{}F(\mathbf{r})\, dV & = \frac{4P\lambda}{hc}\int_{- R}^{R}\frac{1}{w(z)^{2}} \frac{w(z)^{2}}{4}\left\lbrack 1 - \exp\left( - \frac{2a(z)^{2}}{w(z)^{2}} \right) \right\rbrack dz \\
 & = \frac{P\lambda}{hc}\int_{- R}^{R}\left\lbrack 1 - \exp\left( - \frac{2(R^{2} - z^{2})}{w(z)^{2}} \right) \right\rbrack dz.
\end{aligned}
\tag{S15}
\end{equation}
Therefore, the exact droplet-averaged photon flux is,

\begin{equation}
\langle F\rangle = \frac{1}{(4/3)\pi R^{3}}\frac{P\lambda}{hc}\int_{- R}^{R}\left\lbrack 1 - \exp\left( - \frac{2(R^{2} - z^{2})}{w(z)^{2}} \right) \right\rbrack dz
\tag{S16}
\end{equation}
\emph{\textbf{D. Large-droplet limit:}} In the large-droplet limit, the
droplet cross-section is much larger than the beam radius over nearly
the entire path through the droplet,

\begin{equation}
R \gg w(z),
\tag{S17}
\end{equation}
so that,

\begin{equation}
\exp\left( - \frac{2(R^{2} - z^{2})}{w(z)^{2}} \right) \approx 0
\tag{S18}
\end{equation}
for most \(z\). Therefore,

\begin{equation}
1 - \exp\left( - \frac{2(R^{2} - z^{2})}{w(z)^{2}} \right) \approx 1,
\tag{S19}
\end{equation}
and the integral simplifies to,

\begin{equation}
\int_{- R}^{R}1\, dz = 2R.
\tag{S20}
\end{equation}
Substituting into the exact expression yields,

\begin{equation}
\begin{aligned}
\langle F\rangle & \approx \frac{1}{(4/3)\pi R^{3}}\frac{P\lambda}{hc}(2R).
\end{aligned}
\tag{S21}
\end{equation}
Hence, in the large-droplet limit,

\begin{equation}
\langle F\rangle \approx \frac{3}{2\pi}\frac{P\lambda}{hc\, R^{2}}
\tag{S22}
\end{equation}
\emph{\textbf{E. Conversion to photochemical rate constant:}} The
photochemical first-order rate constant is,

\begin{equation}
k = \sigma\langle F\rangle \phi,
\tag{S23}
\end{equation}
where \(\sigma\) is the absorption cross section and \(\phi\) is the
quantum yield. Thus, the exact expression is,

\begin{equation}
k(R) = \frac{\sigma P \lambda \phi}{hc\,(4/3)\pi R^{3}}\int_{- R}^{R}\left\lbrack 1 - \exp\left( - \frac{2(R^{2} - z^{2})}{w(z)^{2}} \right) \right\rbrack dz
\tag{S24}
\end{equation}
and in the large-droplet limit this reduces to,

\begin{equation}
k(R) \approx \frac{3}{2\pi}\frac{\sigma \phi P \lambda}{h c\,R^{2}}
\tag{S25}
\end{equation}
The molar extinction coefficient \(\epsilon\) in units of
\(M^{- 1}cm^{- 1}\) is,

\begin{equation}
\sigma = \frac{1000\ln(10)}{N_{A}}\,\epsilon
\tag{S26}
\end{equation}
where \(N_{A}\) is Avogadro's number. Substituting into the
large-droplet expression yields,

\begin{equation}
k(R) \approx \frac{3}{2\pi}\frac{P \lambda}{h c\, R^{2}}\left( \frac{1000\ln(10)}{N_{A}} \right)\epsilon \phi,
\tag{S27}
\end{equation}
which corresponds to Eq. (1) in the main text.

\section*{S3. Spectra calculation from molecular dynamics simulations}

\textbf{A. Methods}

The computed spectra are obtained from a spin-boson mapping, where the
dynamics on the anharmonic potential-energy surfaces are mapped to
effective harmonic potentials.\cite{ref7,ref8,ref9,ref10,ref11} Within the effective
harmonic model, the vibrational line shape is entirely determined by the
spectral density \(\mathcal{J}(\omega)\), which is directly related to
the spectral distribution
\(\rho(\omega\mathcal{) = J(}\omega)/\pi\Lambda\omega\). Here, the
reorganization energy
\(\Lambda = \frac{\left\langle \delta\Delta V^{2} \right\rangle}{2k_{\text{B}}T}\ \)is
determined by the fluctuations of the energy gap between ground- and
excited-state potentials
\(\delta\Delta V = \Delta V - \left\langle \Delta V \right\rangle_{\text{GS}}\)
relative to its phase-space average in the ground state, with the
vertical energy gap \(\Delta V = V_{\text{ES}} - V_{\text{GS}}\). From
the energy-gap autocorrelation function
\(C(t) = \left\langle \delta\Delta V(t)\delta\Delta V(0) \right\rangle_{\text{GS}}\),
the spectral distribution can be obtained as

\begin{equation}
\rho(\omega) = \frac{2\int_{0}^{\infty}{{C(t)\cos}(\omega t)}dt}{\pi C(0)}
\tag{S28}
\end{equation}
from which the absorption spectrum is calculated as

\begin{equation}
\begin{aligned}
A(\omega) & = \frac{1}{2\pi}\int_{- \infty}^{\infty}e^{i\omega t}c(t)\, dt, \\
c(t) & = \exp\left\lbrack - i\overline{\omega}t - \frac{1}{\pi\hslash}\int_{0}^{\infty}\frac{\mathcal{J}(\omega)}{\omega^{2}}\left( \coth\left( \frac{\hslash\omega}{2k_{\text{B}}T} \right)\left( 1 - \cos(\omega t) \right) + i\left( \sin(\omega t) - \omega t \right) \right)d\omega \right\rbrack,
\end{aligned}
\tag{S29}
\end{equation}
where \(\hslash\bar{\omega}\) is the vertical electronic energy gap of
the effective harmonic model, which sets the absolute position of the
absorption spectrum.

For photoionization, we approximate the continuum of final electron
states by a three-dimensional free-electron density of states
\(\mu(\Omega) \propto \sqrt{\Omega}\,\Theta(\Omega)\), where
\(\hslash\Omega\) is the electron excess energy above the continuum
edge. We then obtain the photoionization spectrum by convolution:

\begin{equation}
\begin{matrix}
A_{\text{ion}}(\omega) = \int_{0}^{\omega}A\left( \omega' \right)\,\mu\left( \omega - \omega' \right)\, d\omega'.
\end{matrix}
\tag{S30}
\end{equation}
Before convolution, the electronic energy gap (\(\hslash\bar{\omega}\))
entering \(A(\omega)\) is lowered by 0.2 eV relative to the
vacuum-electron reference to account approximately for the vertical
electron affinity of bulk water\cite{ref12}.

Numerically, we construct bespoke force fields for nitrite in its ground
state, \(S_{1}\) excited state, and ionized state with the approach
outlined in Ref. \cite{ref7}~ and combine them with the TIP4P-FB water
model.\cite{ref13} The force fields are parametrized from
density-functional theory calculations with the \ensuremath{\omega}B97X functional
combined with D3 dispersion and a polarizable continuum model for water,
employing a ma-def2-QZVPP basis set as implemented in ORCA
6.0.1.\cite{ref14,ref15} We then set up GPU-accelerated molecular-dynamics
simulations as implemented in OpenMM (version 8.1.1)\cite{ref16} of a
nitrite ion in a periodic cubic box of length 27.940~\AA{} with 729 water
molecules, such that the density is 1~g~cm\textsuperscript{-3}. After
thermalization to 300~K with a Langevin thermostat, we initialize 32 NVE
trajectories from different, equilibrated snapshots of the NVT
trajectory, and run the trajectories for 200~ps with a time step of
0.25~fs, from which the energy-gap autocorrelation function is
evaluated.

The absorption intensity of the simulated excitation spectra is scaled
to the experimental spectrum at the absorption peak. The photoionization
band in the experimental spectrum is hidden below the more intense
\(\pi \rightarrow \pi^{*}\) band and thus cannot directly be related to
the simulation result. Instead, we normalize the calculated ionization
spectrum to the effective extinction coefficient
\(\epsilon\Phi = 232\ M^{- 1}cm^{- 1}\) at 193 nm, obtained from the
absorption coefficient and hydrated-electron quantum yield reported by
Sauer \emph{et al.}\cite{ref17} The normalized spectrum therefore already
incorporates the measured electron-production yield.

\textbf{\emph{B. Calibration}}

While the spectral density determines the vibrational line shape, we
further compute the vertical electronic-energy gaps and introduce an
empirical scaling of the intramolecular gap fluctuations to improve
agreement with experiment. We also assess the sensitivity to the water
model and charge-assignment scheme and test the second-order cumulant
approximation.

\emph{B.1 Electronic energy gap}

The electronic energy gap (\(\hslash\bar{\omega}\)) for the
photoexcitation from \emph{S}\textsubscript{0} to
\emph{S}\textsubscript{1} is calculated as the vertical excitation
energy with linear-response time-dependent density functional theory
(LR-TDDFT) from the ground-state minimum of nitrite, using the
electronic-structure setup stated above with equilibrium continuum
solvation for the ground state and ``nonequilibrium'' solvation on the
excited state, excluding longer-timescale solvent reorganization. The
result of \(\hslash{\bar{\omega}}_{exc} = 3.54\ eV\), corresponding to
350 nm, is in reasonable agreement with the experimental band position,
supporting the use of this vertical-energy estimate. For
photoionization, we estimate the vertical detachment energy using \ensuremath{\Delta}SCF
calculations with explicit hydration shells. We select a ground-state MD
configuration whose force-field energy gap is close to the ensemble
average and construct clusters containing nitrite and progressively
larger surrounding solvation shells. For each cluster, we evaluate the
energies of nitrite and the ionized radical using the \ensuremath{\omega}B97X-D3 density
functional and the def2-TZVPP basis set. This leads to the following
results for the vertical electronic gap.

\begin{longtable}[]{@{}
  >{\raggedright\arraybackslash}p{(\linewidth - 4\tabcolsep) * \real{0.1366}}
  >{\centering\arraybackslash}p{(\linewidth - 4\tabcolsep) * \real{0.2587}}
  >{\centering\arraybackslash}p{(\linewidth - 4\tabcolsep) * \real{0.1944}}@{}}
\toprule\noalign{}
\caption{Summary results for energy gaps.}\tabularnewline
\begin{minipage}[b]{\linewidth}\raggedright
Radius {[}\AA{}{]}
\end{minipage} & \begin{minipage}[b]{\linewidth}\raggedright
Nr. of water molecules
\end{minipage} & \begin{minipage}[b]{\linewidth}\raggedright
Energy gap (eV)
\end{minipage} \\
\midrule\noalign{}
\endfirsthead
\toprule\noalign{}
\begin{minipage}[b]{\linewidth}\raggedright
Radius {[}\AA{}{]}
\end{minipage} & \begin{minipage}[b]{\linewidth}\raggedright
Nr. of water molecules
\end{minipage} & \begin{minipage}[b]{\linewidth}\raggedright
Energy gap (eV)
\end{minipage} \\
\midrule\noalign{}
\endhead
\bottomrule\noalign{}
\endlastfoot
7 & 65 & 6.19 \\
9 & 124 & 6.68 \\
11 & 223 & 6.49 \\
\end{longtable}

Given the relatively stable results for the energy gap with varying size
of the solvation shell, we adopt the detachment energy of the largest
cluster. Subtracting the 0.2 eV vertical electron affinity of water
gives \(\hslash{\bar{\omega}}_{ion} = 6.29\ eV\).

\emph{B.2 Comparison of reorganization energies with different water
models and charge schemes}

In order to assess the quality of the employed water model, we present
reorganization energies computed with polarizable and non-polarizable
water models, as well as different schemes to extract point charges from
our electronic structure calculations in Table S1. It can be seen that
the reorganization energies for both excitation and ionization are
rather insensitive to changes in the modelling setup with regard to the
use of polarizable or non-polarizable water models, or the scheme used
to extract point charges from ab initio calculations.

\begin{longtable}[]{@{}
  >{\centering\arraybackslash}p{(\linewidth - 8\tabcolsep) * \real{0.1199}}
  >{\centering\arraybackslash}p{(\linewidth - 8\tabcolsep) * \real{0.1897}}
  >{\centering\arraybackslash}p{(\linewidth - 8\tabcolsep) * \real{0.2077}}
  >{\centering\arraybackslash}p{(\linewidth - 8\tabcolsep) * \real{0.2028}}
  >{\centering\arraybackslash}p{(\linewidth - 8\tabcolsep) * \real{0.2799}}@{}}
\caption{Reorganization energies in kJ/mol with
different non-polarizable TIP4P-FB and polarizable SWM4-NDP water
models, as well as ChelpG and Mulliken charge schemes with and without
continuum polarization (CPCM) for comparison.}\tabularnewline
\toprule\noalign{}
\begin{minipage}[b]{\linewidth}\centering
\end{minipage} & \begin{minipage}[b]{\linewidth}\centering
TIP4P-FB, ChelpG
\end{minipage} & \begin{minipage}[b]{\linewidth}\centering
SWM4-NDP, ChelpG
\end{minipage} & \begin{minipage}[b]{\linewidth}\centering
TIP4P-FB, Mulliken
\end{minipage} & \begin{minipage}[b]{\linewidth}\centering
TIP4P-FB, ChelpG w/o CPCM
\end{minipage} \\
\midrule\noalign{}
\endfirsthead
\toprule\noalign{}
\begin{minipage}[b]{\linewidth}\centering
\end{minipage} & \begin{minipage}[b]{\linewidth}\centering
TIP4P-FB, ChelpG
\end{minipage} & \begin{minipage}[b]{\linewidth}\centering
SWM4-NDP, ChelpG
\end{minipage} & \begin{minipage}[b]{\linewidth}\centering
TIP4P-FB, Mulliken
\end{minipage} & \begin{minipage}[b]{\linewidth}\centering
TIP4P-FB, ChelpG w/o CPCM
\end{minipage} \\
\midrule\noalign{}
\endhead
\bottomrule\noalign{}
\endlastfoot
Excitation & 14.6 & 15.5 & 14.8 & - \\
Ionization & 392 & 360 & - & 410 \\
\end{longtable}

\emph{B.3 Energy-gap analysis and scaling}

In order to assess the accuracy of our fitted force field in capturing
the energy-gap fluctuations, we analyze the contributions to the total
energy gap from intramolecular vibrations of nitrite and intermolecular
electrostatic interactions in Fig. S2. It is shown that the energy-gap
fluctuations in the excitation process are dominated by the
intramolecular nitrite interactions, while the gap fluctuation in the
ionization process is dominated by electrostatic interactions. This is
in line with the intuition that ionization leads to a more substantial
change in the system's charge distribution than the excitation, causing
the electrostatic environment to reorganize more strongly.

\begin{figure}[htbp]
\centering
\includegraphics[width=0.94\linewidth]{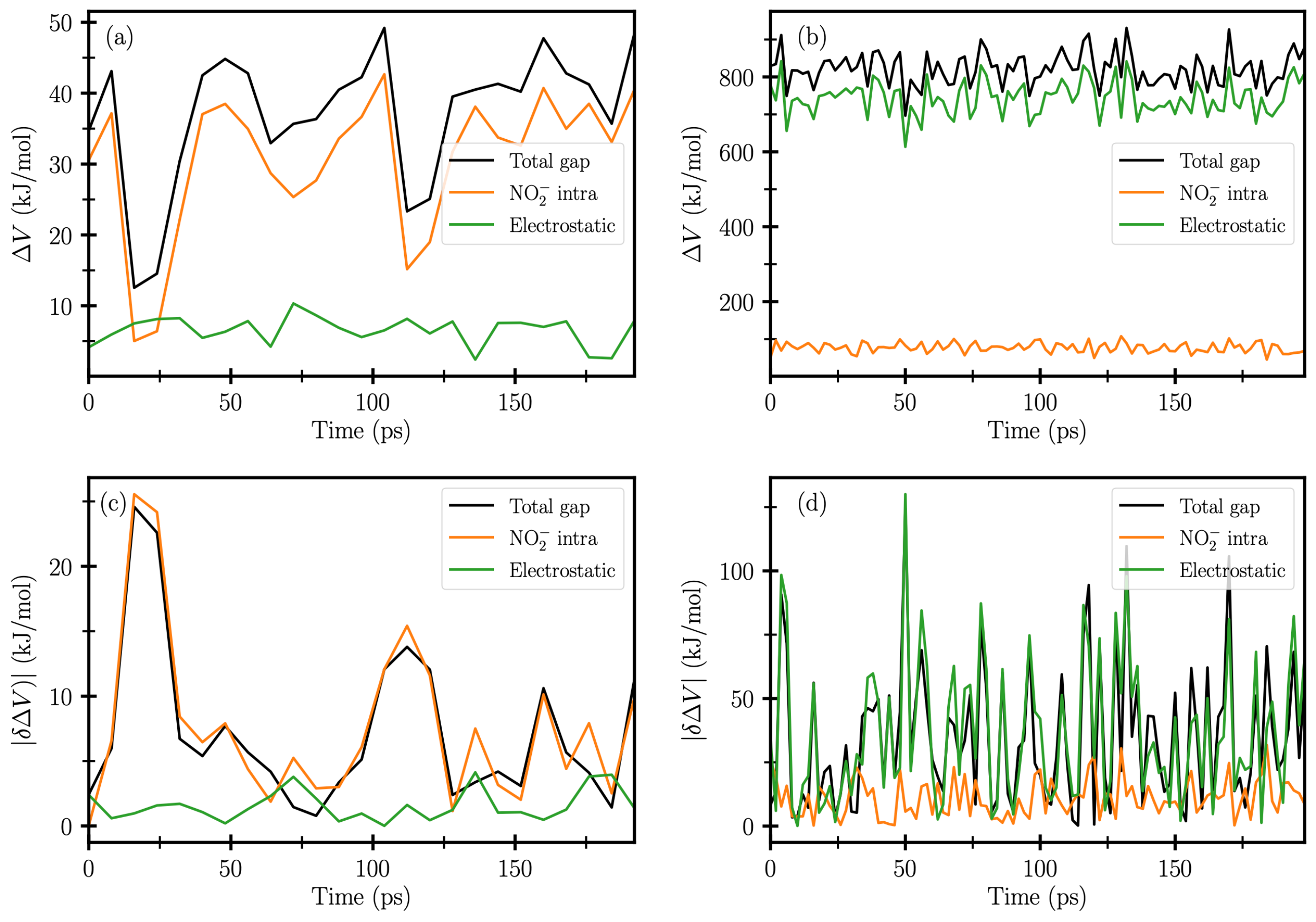}
\caption{Energy gap $\Delta V$ and energy-gap fluctuations $\delta\Delta V$ along a characteristic trajectory for the excitation (a,c) and ionization (b,d) simulation separated into contributions from electrostatic interactions and intramolecular nitrite interactions.}
\label{fig:S2}
\end{figure}

The comparison of spectra in Fig. S3 shows that the calculated
excitation band is narrower than the experimental band. We therefore
introduce an empirical factor of \(a = 2\) in the intramolecular contribution to the energy-gap fluctuations,
\(\delta\Delta V_{a} = a\delta\Delta V_{intra} + \delta\Delta V_{inter}\),
to improve the calculated band shape. This correction is motivated by
limitations of the harmonic intramolecular force field, including its
treatment of anharmonicity. For consistency, the same scaling to the
intramolecular gap is applied to the ionization results in the main
text. Note, however, that the dominance of ionization at long
wavelengths, arising from its broader spectral tail, is independent of
this scaling, which is only meant to provide better quantitative
agreement with experimental intensity scales.

\begin{figure}[htbp]
\centering
\includegraphics[width=0.94\linewidth]{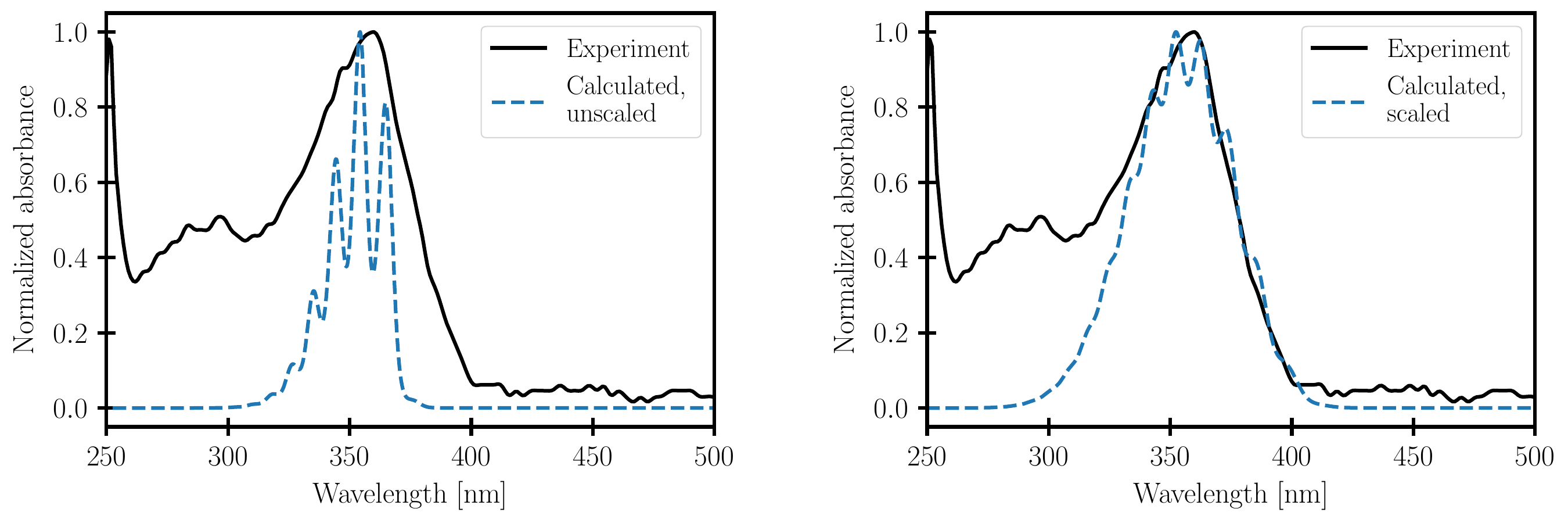}
\caption{Calculated excitation spectrum without scaling (a) and with the factor-2 scaling of the intramolecular nitrite contribution to the energy gap (b), as described in the text, compared to the experimental spectrum. The experimental spectrum contains the onset of a much stronger short-wavelength absorption band, so the comparison is intended only for the illustrated $S_{0} \rightarrow S_{1}$ absorption peak and its long-wavelength tail.}
\label{fig:S3}
\end{figure}

\emph{B.4 Anharmonicity estimate}

\begin{figure}[htbp]
\centering
\includegraphics[width=0.94\linewidth]{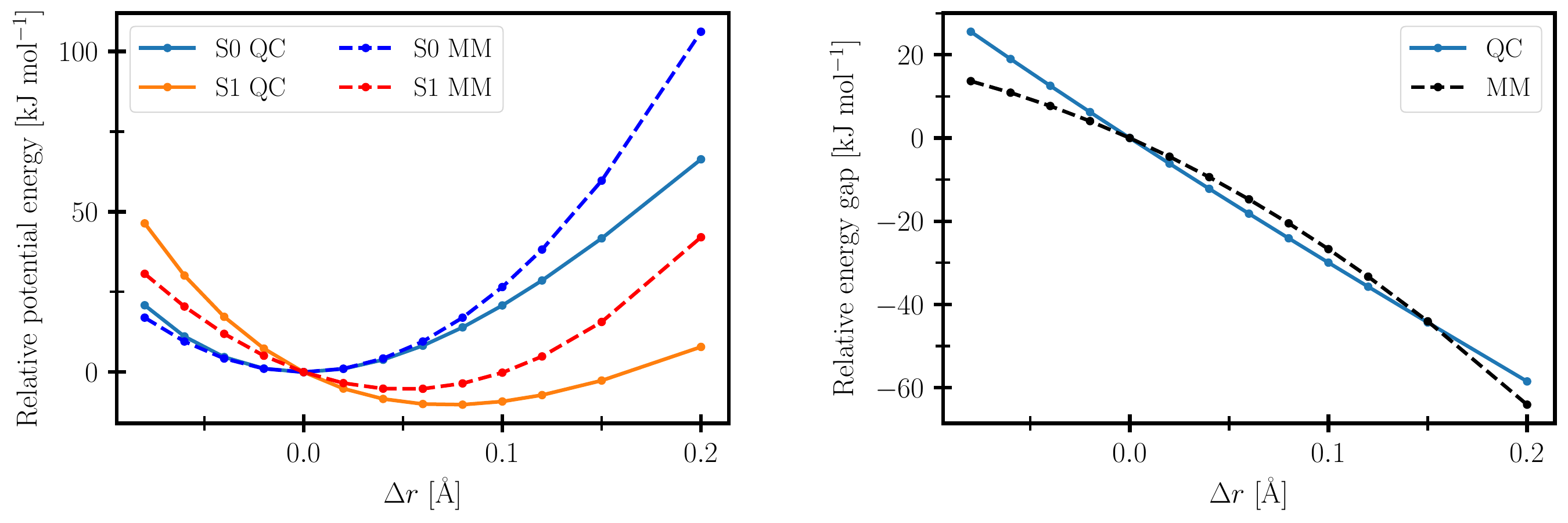}
\caption{Estimate of the effect of anharmonicity on the N--O stretch potential. (Left) Comparison of ab-initio DFT calculations (QC) with our MM force field for the $S_{0}$ and $S_{1}$ energy along displacements $\Delta r$ along a single N--O bond distance relative to the $S_{0}$ minimum geometry with all remaining coordinates fixed. All energies are normalized to a common reference of 0 at zero displacement. (Right) Comparison of the corresponding $S_{0} \rightarrow S_{1}$ energy gaps with 0 displacement as the common reference.}
\label{fig:S4}
\end{figure}

To examine the limitations of the harmonic intramolecular force field,
we compare quantum-chemical and force-field scans of the \(S_{0}\) and
\(S_{1}\) energies of nitrite along an N--O bond (Fig. S4). It can be
seen that the quantum-chemical curves exhibit anharmonicity. We can
formulate an estimate for the correction of the energy-gap fluctuations
of the N--O stretch mode by assuming an approximately harmonic potential
around the ground-state minimum
\(V_{0}(\Delta r) \approx \frac{k}{2}\Delta r^{2}\) and an approximately
linear trend of the energy gap
\(\Delta V(\Delta r) \approx \Delta V(0) + g\Delta r\), where
\(g = \frac{\partial\Delta V}{\partial\Delta r}\) and
\(k = \frac{\partial^{2}V_{0}}{\partial\Delta r^{2}}\). With the
variance of a classical harmonic oscillator
\(\left\langle \Delta r^{2} \right\rangle = \frac{1}{\beta k}\), the gap
standard deviation from the N--O stretch is approximately given by
\(\sigma_{\Delta} = \sqrt{\frac{g^{2}}{\beta k}}\). The ratio between
the quantum-chemical and force-field estimates of this standard
deviation computed from the data in Fig.~S4 is 1.34. The
quantum-chemical scan predicts larger gap fluctuations than the
force-field scan within this local estimate. Although the resulting
factor of 1.34 is smaller than the empirical factor of 2, it supports
the direction of the correction and indicates that the force field
underestimates the intramolecular contribution to the gap fluctuations.

\emph{B.5 Correlation functions and energy-gap statistics}

As described in Sec. S3.A, we compute absorption line shapes using a
second-order cumulant treatment, with nuclear quantum effects
incorporated approximately through the effective harmonic mapping.
Examples of energy-gap autocorrelation functions for nitrite ionization
and excitation used in this work are shown in Fig. S5.

\begin{figure}[t]
\centering
\includegraphics[width=0.88\linewidth]{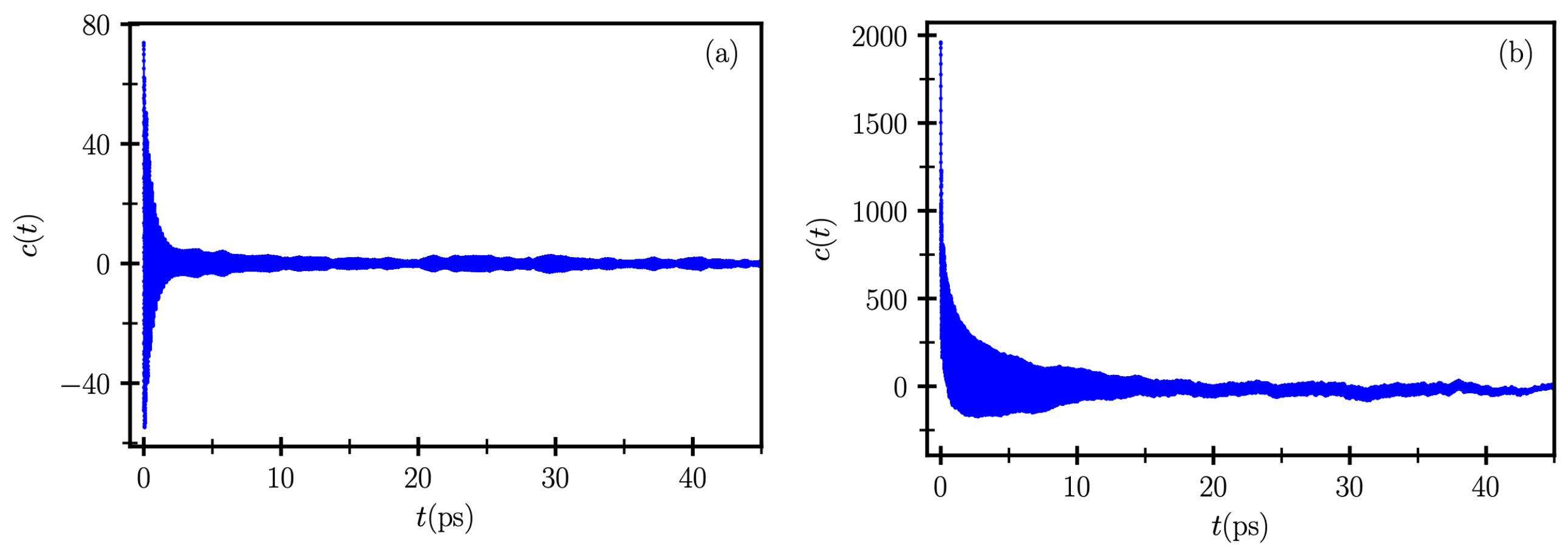}
\caption{Energy gap autocorrelation functions for excitation (a) and ionization (b).}
\label{fig:S5}
\end{figure}

To assess the second-order cumulant approximation, we first compare the
energy-gap distributions sampled from the ground-state MD trajectories
with Gaussian fits (Fig. S6). The ionization gap is approximately
Gaussian distributed, whereas the excitation gap exhibits more
pronounced deviations, particularly in the distribution tails.

\begin{figure}[htbp]
\centering
\includegraphics[width=0.88\linewidth]{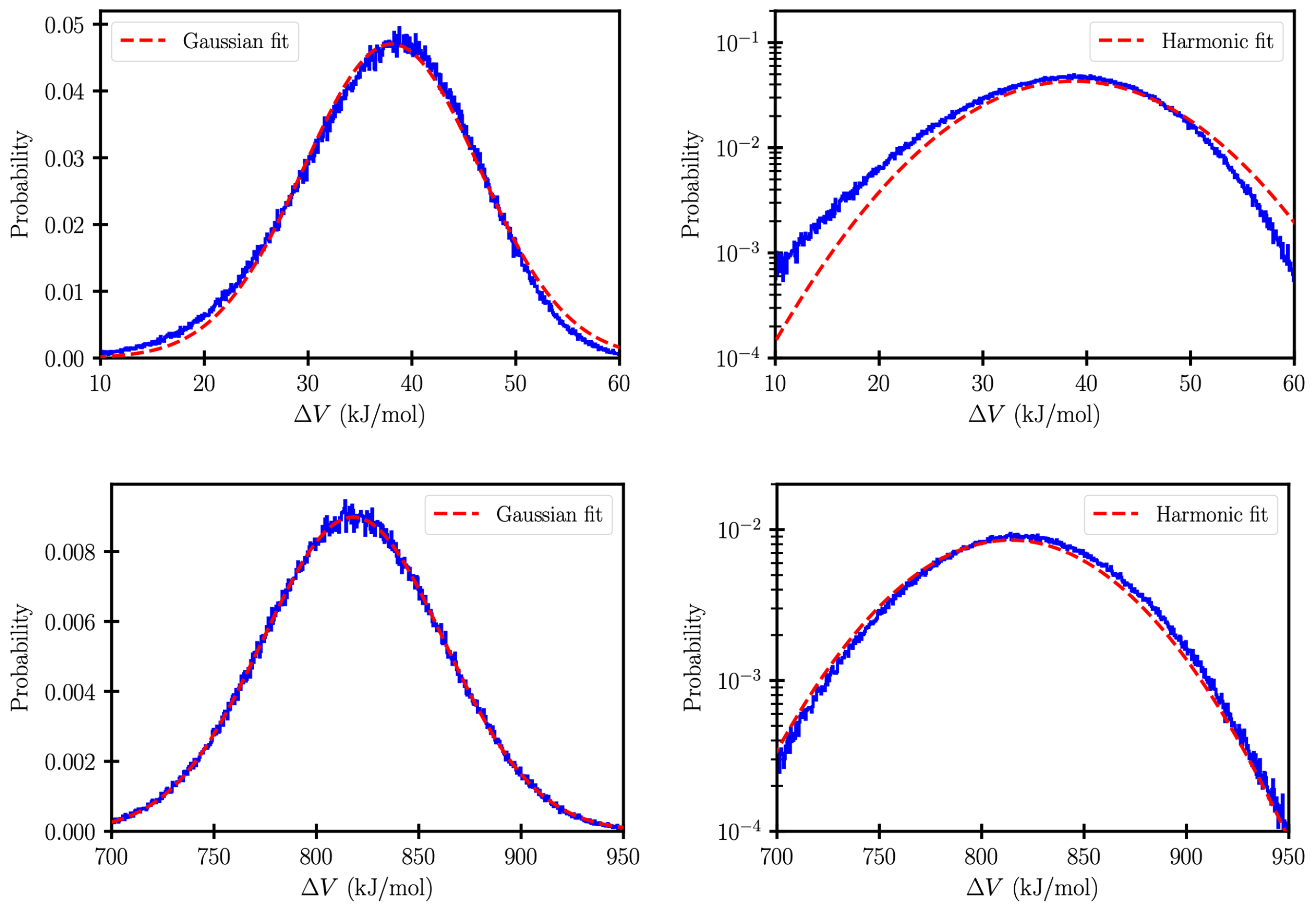}
\caption{Energy gap distribution on linear (left) and logarithmic (right) y-scales for the excitation (top) and ionization (bottom) with corresponding Gaussian and harmonic fits.}
\label{fig:S6}
\end{figure}

We then
test the approximation directly by comparing the cumulant spectra with
classical spectra calculated from the MD trajectories without truncating
the cumulant expansion. For this calculation, we integrate the
energy-gap fluctuations over each trajectory segment to obtain the
accumulated phase. Averaging the corresponding complex exponentials over
trajectories and starting times gives

\[c_{full}^{cl}(t) = e^{- i\overline{\omega}t}\left\langle \exp\left\lbrack - \frac{i}{\hslash}\int_{0}^{t}{\delta\Delta V\left( t_{0} + s \right)\ ds} \right\rbrack \right\rangle_{GS,t_{0}}.\]

\begin{figure}[htbp]
\centering
\includegraphics[width=0.88\linewidth]{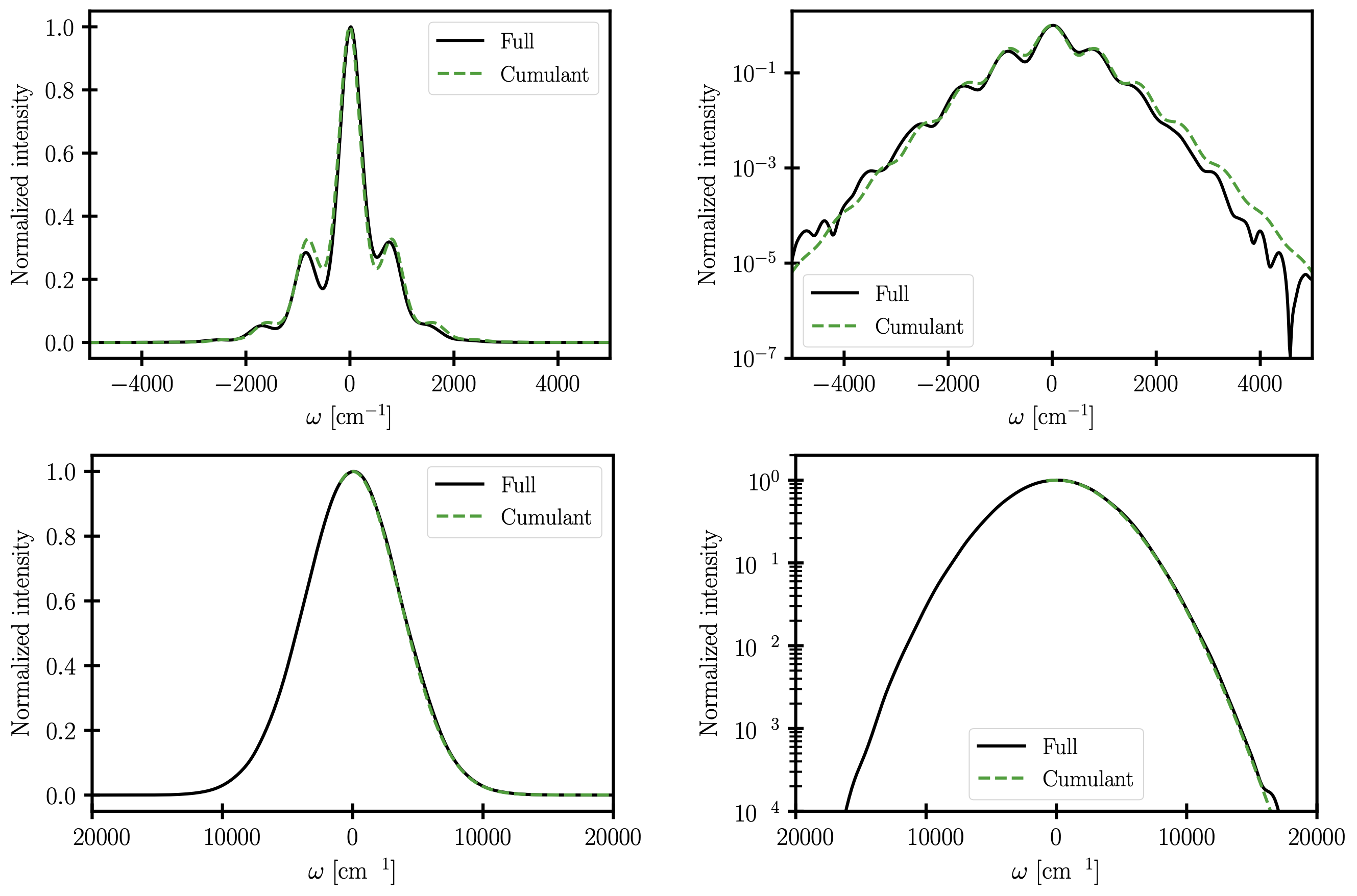}
\caption{Comparison of classical vibrational line shapes as a function of the wavenumber offset from the absorption peak obtained by direct phase averaging (``Full'') and the second-order cumulant approximation on linear (left) and logarithmic (right) y-scales for the excitation (top) and ionization (bottom).}
\label{fig:S7}
\end{figure}

Fourier transformation yields the direct classical spectra shown in Fig.
S7. These agree closely with the second-order cumulant spectra for
ionization and show good agreement for excitation, with somewhat larger
differences in the excitation-band tails. Overall, this comparison
supports the second-order cumulant approximation over the resolved
spectral range.

\textbf{\emph{C. Photochemical rate}}

From the excitation and ionization spectra, corresponding photochemical
rates under solar irradiation can be computed. The photochemical rates
for excitation and ionization are given by
\(J = \int\sigma(\lambda)\Phi(\lambda)F(\lambda)\, d\lambda\), with the
respective absorption cross section \(\sigma\), the quantum yield for
the process of interest \(\Phi\), and the spectral actinic flux \(F\).
For photoionization, the product \(\sigma(\lambda)\Phi(\lambda)\) is
supplied directly by the normalized effective spectrum. Here, the solar
photon flux \(F\) was derived from the ASTM G-173 AM1.5G global-tilted
reference spectrum distributed by NREL,\cite{ref18} converted from
spectral irradiance \(E(\lambda)\) to photons cm\(^{- 2}\) s\(^{- 1}\)
nm\(^{- 1}\) through
\(F(\lambda) = E(\lambda)\frac{\lambda}{hc}10^{- 4}\). The cross section
is taken from the computed extinction coefficient
\(\sigma(\lambda) = \epsilon(\lambda)\frac{\ln 10}{N_{\text{A}}}10^{3}\),
with Avogadro constant \(N_{\text{A}}\).

The results are reported in Table S2 with and without the empirical
scaling of the intramolecular energy-gap fluctuations. It can be seen
that the excitation rate is not very sensitive to the scaling because
the rate is dominated by the center of the absorption peak. The
ionization rate on the other hand is dominated by the long-wavelength
tail, rendering the rate estimate highly sensitive to scaling and
inaccuracies in the energy gaps. The qualitative enhancement of
ionization relative to excitation at long wavelengths is therefore more
robust than its absolute photochemical rate.

\begin{longtable}[]{@{}
  >{\raggedright\arraybackslash}p{(\linewidth - 8\tabcolsep) * \real{0.1133}}
  >{\centering\arraybackslash}p{(\linewidth - 8\tabcolsep) * \real{0.1291}}
  >{\centering\arraybackslash}p{(\linewidth - 8\tabcolsep) * \real{0.3357}}
  >{\centering\arraybackslash}p{(\linewidth - 8\tabcolsep) * \real{0.1277}}
  >{\centering\arraybackslash}p{(\linewidth - 8\tabcolsep) * \real{0.2502}}@{}}
\caption{Nitrite photoionization and photoexcitation
rates \(J\) in \(s^{- 1}\) from spectra with and without the empirical
scaling of the energy gap. The excitation rate is reported both before
and after applying a 2\% quantum yield for OH radical production.\cite{ref19,ref20,ref21,ref22,ref23}
The ionization result implicitly contains the quantum yield for events
producing free hydrated electrons surviving the early recombination
dynamics because the intensity is normalized by the experimental
hydrated-electron quantum yield at 193~nm.\cite{ref17} For comparison, we
compute the ionization rate when the effective extinction is normalized
to our experimentally inferred effective extinction coefficient
\(\epsilon \Phi = 3.46 \times 10^{- 6}\ M^{- 1}cm^{- 1}\) at 532
nm.}\tabularnewline
\toprule\noalign{}
\begin{minipage}[b]{\linewidth}\raggedright
\end{minipage} & \begin{minipage}[b]{\linewidth}\raggedright
Excitation
\end{minipage} & \begin{minipage}[b]{\linewidth}\raggedright
Excitation with quantum yield
\end{minipage} & \begin{minipage}[b]{\linewidth}\raggedright
Ionization
\end{minipage} & \begin{minipage}[b]{\linewidth}\raggedright
Ionization normalized\\
at 532 nm\strut
\end{minipage} \\
\midrule\noalign{}
\endfirsthead
\toprule\noalign{}
\begin{minipage}[b]{\linewidth}\raggedright
\end{minipage} & \begin{minipage}[b]{\linewidth}\raggedright
Excitation
\end{minipage} & \begin{minipage}[b]{\linewidth}\raggedright
Excitation with quantum yield
\end{minipage} & \begin{minipage}[b]{\linewidth}\raggedright
Ionization
\end{minipage} & \begin{minipage}[b]{\linewidth}\raggedright
Ionization normalized\\
at 532 nm\strut
\end{minipage} \\
\midrule\noalign{}
\endhead
\bottomrule\noalign{}
\endlastfoot
unscaled & \(2 \times 10^{- 4}\) & \(3 \times 10^{- 6}\) &
\(3 \times 10^{- 9}\) & \(-\) \\
scaled & \(5 \times 10^{- 4}\) & \(1 \times 10^{- 5}\) &
\(9 \times 10^{- 7}\) & \(2 \times 10^{- 6}\) \\
\end{longtable}

\begin{figure}[htbp]
\centering
\includegraphics[width=0.94\linewidth]{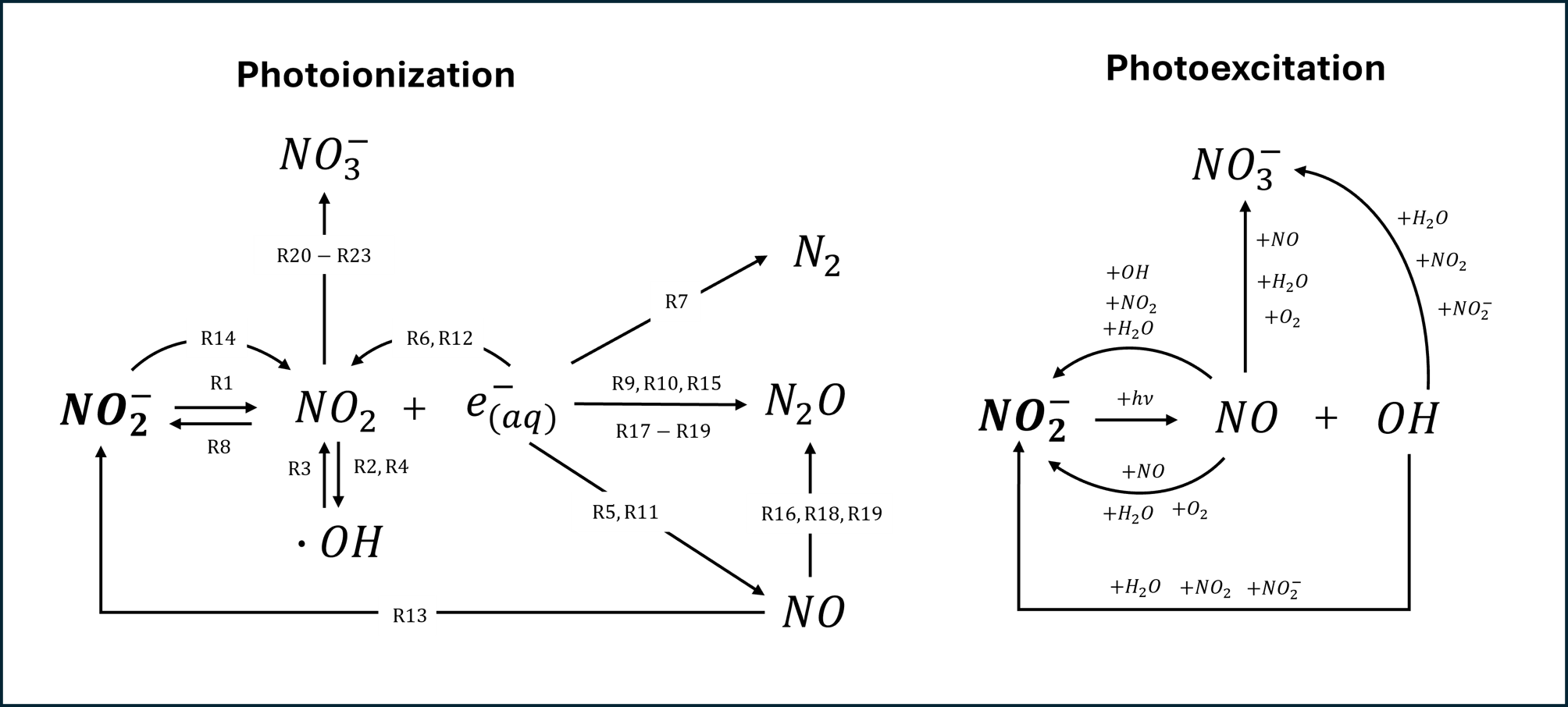}
\caption{Skeletal mechanisms for photoionization and photoexcitation. Photoionization pathways are labeled with the explicit steps in the mechanism shown in Table S3.}
\label{fig:S8}
\end{figure}

\section*{S4. Photoionization and Photoexcitation Mechanisms}

\textbf{Table 4:} Photoionization mechanism, rate and extinction
coefficients with literature references.

{\def\LTcaptype{none} % do not increment counter
\begin{longtable}[]{@{}
  >{\centering\arraybackslash}p{(\linewidth - 6\tabcolsep) * \real{0.0720}}
  >{\centering\arraybackslash}p{(\linewidth - 6\tabcolsep) * \real{0.3991}}
  >{\centering\arraybackslash}p{(\linewidth - 6\tabcolsep) * \real{0.4043}}
  >{\raggedright\arraybackslash}p{(\linewidth - 6\tabcolsep) * \real{0.1246}}@{}}
\toprule\noalign{}
\begin{minipage}[b]{\linewidth}\centering
\#
\end{minipage} & \begin{minipage}[b]{\linewidth}\centering
Reaction
\end{minipage} & \begin{minipage}[b]{\linewidth}\centering
Rate/Extinction Coeff.
\end{minipage} & \begin{minipage}[b]{\linewidth}\centering
Ref.
\end{minipage} \\
\midrule\noalign{}
\endhead
\bottomrule\noalign{}
\endlastfoot
& \textbf{Photochemistry} & & \\
R1 &
\({NO}_{2}^{-}\overset{h\nu}{\rightarrow}\ {NO}_{2} + e_{(aq)}^{-}\) &
See Eq. (1)

\ensuremath{\epsilon\phi} = (3.46 \ensuremath{\pm} 0.19) x 10\textsuperscript{-6} M\textsuperscript{-1}
cm\textsuperscript{-1} & This work \\
R2 & \({NO}_{2}\overset{h\nu}{\rightarrow}\ {NO}_{2}^{*}\) & \ensuremath{\epsilon} = 38
M\textsuperscript{-1} cm\textsuperscript{-1} & a \\
R3 & \({NO}_{2}^{*} + \ H_{2}O \rightarrow \ {NO}_{2} + \ H_{2}O\) &
\(1.02\  \times \ 10^{11}\ M^{- 1}s^{- 1}\) & b \\
R4 &
\({NO}_{2}^{*} + \ H_{2}O \rightarrow \ {NO}_{2}^{-} + \ {OH}^{\bullet}\)
+ \(H^{+}\) &
\(\left( 4.5\  \times \ 10^{5} - 1.02 \times \ 10^{8} \right)\ M^{- 1}s^{- 1}\)
& c \\
& & & \\
& \textbf{Intermediate Pathways} & & \\
R5 & \({NO}_{2}^{-} + e_{(aq)}^{-} \rightarrow {NO}_{2}^{2 -}\) &
\(4.1\  \times \ 10^{9}\ M^{- 1}s^{- 1}\) & d \\
R6 & \({NO}_{3}^{-} + e_{(aq)}^{-} \rightarrow {NO}_{3}^{2 -}\) &
\(9.7\  \times \ 10^{9}\ M^{- 1}s^{- 1}\) & d \\
R7 &
\(N_{2}O + e_{(aq)}^{-} + \ H_{2}O \rightarrow N_{2} + \ {OH}^{\bullet} + {OH}^{-}\)
& \(9.1\  \times \ 10^{9}\ M^{- 1}s^{- 1}\) & d \\
R8 & \({NO}_{2} + e_{(aq)}^{-} \rightarrow {NO}_{2}^{-}\) &
\(4.6\  \times \ 10^{10}\ M^{- 1}s^{- 1}\) & e \\
R9 & \(NO + e_{(aq)}^{-} \rightarrow {{NO}}^{-}\) &
\(2.3\  \times \ 10^{10}\ M^{- 1}s^{- 1}\) & d \\
R10 &
\({{NO}}^{-} + \ H_{2}O \rightarrow \ {HNO + \ {OH}^{-}}\)
&
\(k\left\lbrack H_{2}O \right\rbrack = 1.2\  \times \ 10^{2}\ s^{- 1}\)
& f \\
R11 & \({NO}_{2}^{2 -} + \ H_{2}O \rightarrow \ NO + 2{OH}^{-}\) &
\(9\  \times \ 10^{4}\ M^{- 1}s^{- 1}\) & e \\
R12 & \({NO}_{3}^{2 -} + \ H_{2}O \rightarrow\) \({NO}_{2} + 2{OH}^{-}\)
&
\(k\left\lbrack H_{2}O \right\rbrack = 5.5\  \times \ 10^{4}\ s^{- 1}\)
& e \\
R13 & \(NO + {OH}^{\bullet} \rightarrow {NO}_{2}^{-} + \ H^{+}\) &
\(1\  \times \ 10^{10}\ M^{- 1}s^{- 1}\) & g \\
R14 & \({NO}_{2}^{-} + {OH}^{\bullet}\  \rightarrow\)
\({NO}_{2} + \ {OH}^{-}\) & \(1\  \times \ 10^{10}\ M^{- 1}s^{- 1}\) &
g \\
R15 & \(2HNO \rightarrow N_{2}O +\) \(H_{2}O\) &
\(8\  \times \ 10^{6}\ M^{- 1}s^{- 1}\) & f \\
R16 & \(HNO + \ NO \rightarrow N_{2}O_{2}^{-} + \ H^{+}\) &
\(5.8\  \times \ 10^{6}\ M^{- 1}s^{- 1}\) & f \\
R17 & \({NO}^{-} + NO\  \rightarrow N_{2}O_{2}^{-}\) &
\(2.3\  \times \ 10^{9}\ M^{- 1}s^{- 1}\) & f \\
R18 & \(N_{2}O_{2}^{-} + NO\  \rightarrow \ N_{3}O_{3}^{-}\) &
\(2.3\  \times \ 10^{9}\ M^{- 1}s^{- 1}\) & f \\
R19 & \(N_{3}O_{3}^{-} \rightarrow N_{2}O + {NO}_{2}^{-}\) &
\(3\  \times \ 10^{2}\ s^{- 1}\) & h \\
R20 & \({2NO}_{2} \rightarrow \ N_{2}O_{4}\) &
\(4.5\  \times \ 10^{8}\ M^{- 1}s^{- 1}\) & e \\
R21 & \(N_{2}O_{4} \rightarrow {2NO}_{2}\) &
\(6.9\  \times \ 10^{3}\ s^{- 1}\) & e \\
& & & \\
& \textbf{Nitrate Formation} & & \\
R22 & \(N_{2}O_{4} + \ H_{2}O \rightarrow {NO}_{3}^{-} + {NO}_{2}^{-}\)
+ 2\(H^{+}\) &
\(k\left\lbrack H_{2}O \right\rbrack = 1\  \times \ 10^{3}\ s^{- 1}\) &
e \\
R23 & \({NO}_{2} + {OH}^{\bullet} \rightarrow {NO}_{3}^{-} + H^{+}\) &
\(1.3\  \times \ 10^{9}\ M^{- 1}s^{- 1}\) & e \\
& & & \\
& \textbf{Scavenger Reactions} & & \\
R24 & \(Fumerate + e_{(aq)}^{-} \rightarrow Prod\_ 1\) &
\(2.2\  \times \ 10^{10}\ M^{- 1}s^{- 1}\) & i \\
R25 & \(Fumerate + {OH}^{\bullet} \rightarrow Prod\_ 2\) &
\(5\  \times \ 10^{9}\ M^{- 1}s^{- 1}\) & i \\
R26 & \({SCN}^{-} + e_{(aq)}^{-} \rightarrow Prod\_ 3\) &
\(< 1\  \times \ 10^{6}\ M^{- 1}s^{- 1}\) & d \\
R27 &
\({SCN}^{-} + {OH}^{\bullet} \rightarrow \ldots \rightarrow {SO}_{4}^{2 -}\)
& \(1.4\  \times \ 10^{10}\ M^{- 1}s^{- 1}\) & j \\
\end{longtable}
}

\textsuperscript{a} Ref. \cite{ref24}, \textsuperscript{b} Ref.
\cite{ref25}, \textsuperscript{c}Ref. \cite{ref25,ref26,ref27,ref28},
\textsuperscript{d}Ref. \cite{ref29}, \textsuperscript{e}Ref.
\cite{ref30}, \textsuperscript{f} Ref. \cite{ref31},
\textsuperscript{g}Ref. \cite{ref21}, \textsuperscript{h}Ref.
\cite{ref32}, \textsuperscript{i}Ref. \cite{ref33},
\textsuperscript{j}Ref. \cite{ref34}

The mechanism shown in Table S3 is drawn from previous literature and
includes a reasonable set of mechanistic steps based upon previous
pulsed radiolysis studies of solvated electron chemistry. The
simulations of our experimental conditions are found to be insensitive
to the range of values reported for R4. To model our experimental
conditions, we do not include reactions steps corresponding to the
formation and equilibrium of N\textsubscript{2}O\textsubscript{3}
produced via the NO + NO\textsubscript{2} reaction. These steps are
omitted, since there is evidence\cite{ref35,ref36} that once
N\textsubscript{2}O\textsubscript{3} is formed, it can undergo rapid
photoisomerization at low laser power at visible wavelengths
(\emph{i.e.,} 632 nm), which is accompanied, for certain isomers, by
dissociation back into NO + NO\textsubscript{2}. Given this previous
literature, we assume that if any N\textsubscript{2}O\textsubscript{3}
is formed in our droplet experiments it will rapidly undergo
photoisomerization/dissociation back into NO and NO\textsubscript{2}.
For simulations under atmospheric and environmental conditions, however,
these N\textsubscript{2}O\textsubscript{3} pathways are included for
both the photoionization (Table S4, R28 and R29) and photoexcitation
(Table S5, E4 and E5) mechanisms.

We also don't include solvated electrons reactions with
O\textsubscript{2}, since under our experimental conditions (2.5 M
Nitrite) the main sink for these species is nitrite and nitrate. This is
not the case however, under atmospheric conditions where nitrite and
other salt concentrations are highly dilute (nM) and the main sink for
solvated electrons is likely dissolved O\textsubscript{2}. The reaction
steps added to the model to simulate these dilute
environmental/atmospheric conditions are shown in Table S4

\textbf{Table 5:} Elementary reaction steps added to the mechanism
shown in Table S3 to simulate photoionization under dilute
atmospheric/environmental conditions. Rate constants and reaction steps
R30-R42 are from Buxton et al.\cite{ref29} The dissolved
O\textsubscript{2} concentration is computed from the partial pressure
of O\textsubscript{2} in the atmosphere and the Henry's law constant for
O\textsubscript{2} in pure water (H = 1.32 x 10\textsuperscript{-3}
M/atm).\cite{ref37} Literature references for R28 and R29 are shown in
Table S5 (E4 and E5).

{\def\LTcaptype{none} % do not increment counter
\begin{longtable}[]{@{}
  >{\centering\arraybackslash}p{(\linewidth - 4\tabcolsep) * \real{0.0862}}
  >{\raggedright\arraybackslash}p{(\linewidth - 4\tabcolsep) * \real{0.4142}}
  >{\raggedright\arraybackslash}p{(\linewidth - 4\tabcolsep) * \real{0.3470}}@{}}
\toprule\noalign{}
\begin{minipage}[b]{\linewidth}\centering
\#
\end{minipage} & \begin{minipage}[b]{\linewidth}\centering
Reaction
\end{minipage} & \begin{minipage}[b]{\linewidth}\centering
Rate/Extinction Coefficient
\end{minipage} \\
\midrule\noalign{}
\endhead
\bottomrule\noalign{}
\endlastfoot
R28 & \(NO\  + \ {NO}_{2} \rightarrow \ N_{2}O_{3}\) &
\(1.0 \times 10^{9}{\ M}^{- 1}s^{- 1}\) \\
R29 & \(N_{2}O_{3} + H_{2}O \rightarrow \ 2H^{+ \ } + 2{NO}_{2}^{-}\) &
\(k\left\lbrack H_{2}O \right\rbrack = 5.3 \times 10^{2}\ s^{- 1}\) \\
R30 & \(O_{2} + e_{(aq)}^{-} \rightarrow O_{2}^{\bullet \  -}\) &
\(k\left\lbrack O_{2} \right\rbrack = 5.26 \times 10^{6}\ s^{- 1}\) \\
R31 &
\(2\ e_{(aq)}^{-} + 2\ H_{2}O\  \rightarrow \ H_{2} + {2\ OH}^{-}\) &
\(0.5 \times 10^{10}{\ M}^{- 1}s^{- 1}\) \\
R32 & \(e_{(aq)}^{-} + \ H^{+ \ }\  \rightarrow H^{\bullet \ }\ \ \) &
\(2.3 \times 10^{10}{\ M}^{- 1}s^{- 1}\) \\
R33 &
\(e_{(aq)}^{-} + \ H^{\bullet \ } + H_{2}O \rightarrow \ H_{2} + {OH}^{-}\)
& \(2.5 \times 10^{10}{\ M}^{- 1}s^{- 1}\) \\
R34 & \({2\ H}^{\bullet \ } \rightarrow \ H_{2}\) &
\(0.75 \times 10^{10}{\ M}^{- 1}s^{- 1}\) \\
R35 &
\(e_{(aq)}^{-} + H_{2}O \rightarrow \ H^{\bullet \ } + {\ OH}^{-}\) &
\(k\left\lbrack H_{2}O \right\rbrack = 1.045 \times 10^{3}\ s^{- 1}\) \\
R36 & \(e_{(aq)}^{-} + {\ OH}^{\bullet} \rightarrow {\ OH}^{-}\) &
\(3 \times 10^{10}{\ M}^{- 1}s^{- 1}\) \\
R37 &
\(\ H^{\bullet \ } + {\ OH}^{-} \rightarrow \ e_{(aq)}^{-} + \ H_{2}O\)
& \(2.2 \times 10^{7}{\ M}^{- 1}s^{- 1}\) \\
R38 & \(2\ {\ OH}^{\bullet} \rightarrow \ H_{2}O_{2}\) &
\(0.55 \times 10^{10}{\ M}^{- 1}s^{- 1}\ \) \\
R39 &
\(H^{\bullet \ } + \ \ H_{2}O_{2} \rightarrow {\ OH}^{\bullet} + H_{2}O\)
& \(9\  \times \ 10^{7}\ M^{- 1}s^{- 1}\) \\
R40 & \(H^{\bullet \ } + {\ OH}^{\bullet} \rightarrow H_{2}O\) &
\(7\  \times \ 10^{9}\ M^{- 1}s^{- 1}\) \\
R41 & \(H_{2} + {\ OH}^{\bullet} \rightarrow \ H^{\bullet \ } + H_{2}O\)
& \(4.2\  \times \ 10^{7}\ M^{- 1}s^{- 1}\) \\
R42 &
\(e_{(aq)}^{-} + H_{2}O_{2} \rightarrow {\ OH}^{\bullet} + {\ OH}^{-}\)
& \(1.0 \times 10^{10}{\ M}^{- 1}s^{- 1}\) \\
\end{longtable}
}

\textbf{Table 6:} Photoexcitation mechanism with literature references.

{\def\LTcaptype{none} % do not increment counter
\begin{longtable}[]{@{}
  >{\centering\arraybackslash}p{(\linewidth - 6\tabcolsep) * \real{0.0861}}
  >{\raggedright\arraybackslash}p{(\linewidth - 6\tabcolsep) * \real{0.4138}}
  >{\raggedright\arraybackslash}p{(\linewidth - 6\tabcolsep) * \real{0.3466}}
  >{\raggedright\arraybackslash}p{(\linewidth - 6\tabcolsep) * \real{0.1535}}@{}}
\toprule\noalign{}
\begin{minipage}[b]{\linewidth}\centering
\#
\end{minipage} & \begin{minipage}[b]{\linewidth}\centering
Reaction
\end{minipage} & \begin{minipage}[b]{\linewidth}\centering
Rate Coefficient
\end{minipage} & \begin{minipage}[b]{\linewidth}\centering
Ref.
\end{minipage} \\
\midrule\noalign{}
\endhead
\bottomrule\noalign{}
\endlastfoot
E1 & \({NO}_{2}^{-}\overset{h\nu}{\rightarrow}\ NO\  + \ OH \bullet\) &
\(J\  = 1.0\  \times 10^{- 5\ }s^{- 1}\) & This Work \\
E2 & \(NO + \ OH \  \rightarrow \ HNO_{2}\) &
\(1.0 \times 10^{10}{\ M}^{- 1}s^{- 1}\) & \textsuperscript{A} \\
E3 & \(OH + \ {NO}_{2}^{-} \rightarrow \ {NO}_{2} + {OH}^{-}\ \)
& \(1.0 \times 10^{10}{\ M}^{- 1}s^{- 1}\) & \textsuperscript{A} \\
E4 & \(NO\  + \ {NO}_{2} \rightarrow \ N_{2}O_{3}\) &
\(1.0 \times 10^{9}{\ M}^{- 1}s^{- 1}\) & \textsuperscript{A} \\
E5 & \(N_{2}O_{3} + H_{2}O \rightarrow \ 2H^{+ \ } + 2{NO}_{2}^{-}\) &
\(k\left\lbrack H_{2}O \right\rbrack = 5.3 \times 10^{2}\ s^{- 1}\) &
\textsuperscript{A} \\
E6 & \({2NO}_{2} \rightarrow N_{2}O_{4}\) &
\(4.5 \times 10^{8}{\ M}^{- 1}s^{- 1}\) & \textsuperscript{A} \\
E7 & \(N_{2}O_{4} \rightarrow {2NO}_{2}\) &
\(6.9\  \times \ 10^{3}\ s^{- 1}\) & \textsuperscript{B} \\
E8 &
\(N_{2}O_{4} + \ H_{2}O \rightarrow {NO}_{2}^{-} + {NO}_{3}^{-} + 2H^{+ \ }\)
& \(k\left\lbrack H_{2}O \right\rbrack = 1.0 \times 10^{3}\ s^{- 1}\) &
\textsuperscript{A} \\
E9 & \(2\ NO\  + \ O_{2}\  \rightarrow {2NO}_{2}\) &
\(k\left\lbrack O_{2} \right\rbrack = \ 567\ {\ M}^{- 1}s^{- 1}\ \) &
\textsuperscript{C} \\
E10 &
\({NO}_{2}\  + \ OH \  \rightarrow {NO}_{3}^{-}\  + H^{+ \ }\) &
\(1.3\  \times \ 10^{9}\ M^{- 1}s^{- 1}\) & \textsuperscript{B} \\
\end{longtable}
}

\textsuperscript{A} Ref. \cite{ref21}, \textsuperscript{B} Ref.
\cite{ref30}, \textsuperscript{C} Ref. \cite{ref38}

\section*{S5. Photoionization Mechanism Validation}

After verifying that both scavengers are photostable in aqueous
droplets, we conducted a series of measurements with aqueous nitrite
(2.38-2.5 M) droplets doped with either 0.72 M fumarate or 2.51 M
SCN\textsuperscript{-}. Shown in Fig. S9 are the time-dependent Raman
spectra and decay kinetics of these scavengers. The Raman spectra in
Fig. S9A shows that the \({SCN}^{-}\) peak at 2000
cm\textsuperscript{-1} decays over time as does the peak corresponding
to nitrite. Nitrate (1055 cm\textsuperscript{-1}) is observed to be
formed over the course the reaction as well as there is evidence for
sulfate formation (981 cm\textsuperscript{-1}); a terminal product,
formed after multiple oxidation steps of \({SCN}^{-}\). The decay
kinetics for \({SCN}^{-}\) are shown in Fig. S9B. The kinetic model
replicates the \({SCN}^{-}\) decay kinetics as well as the overall
formation of sulfate. Since \({SCN}^{-}\) is a selective scavenger for
OH, this confirms that the OH production pathways in the kinetic model
are reasonably accurate.

\begin{figure}[htbp]
\centering
\includegraphics[width=0.94\linewidth]{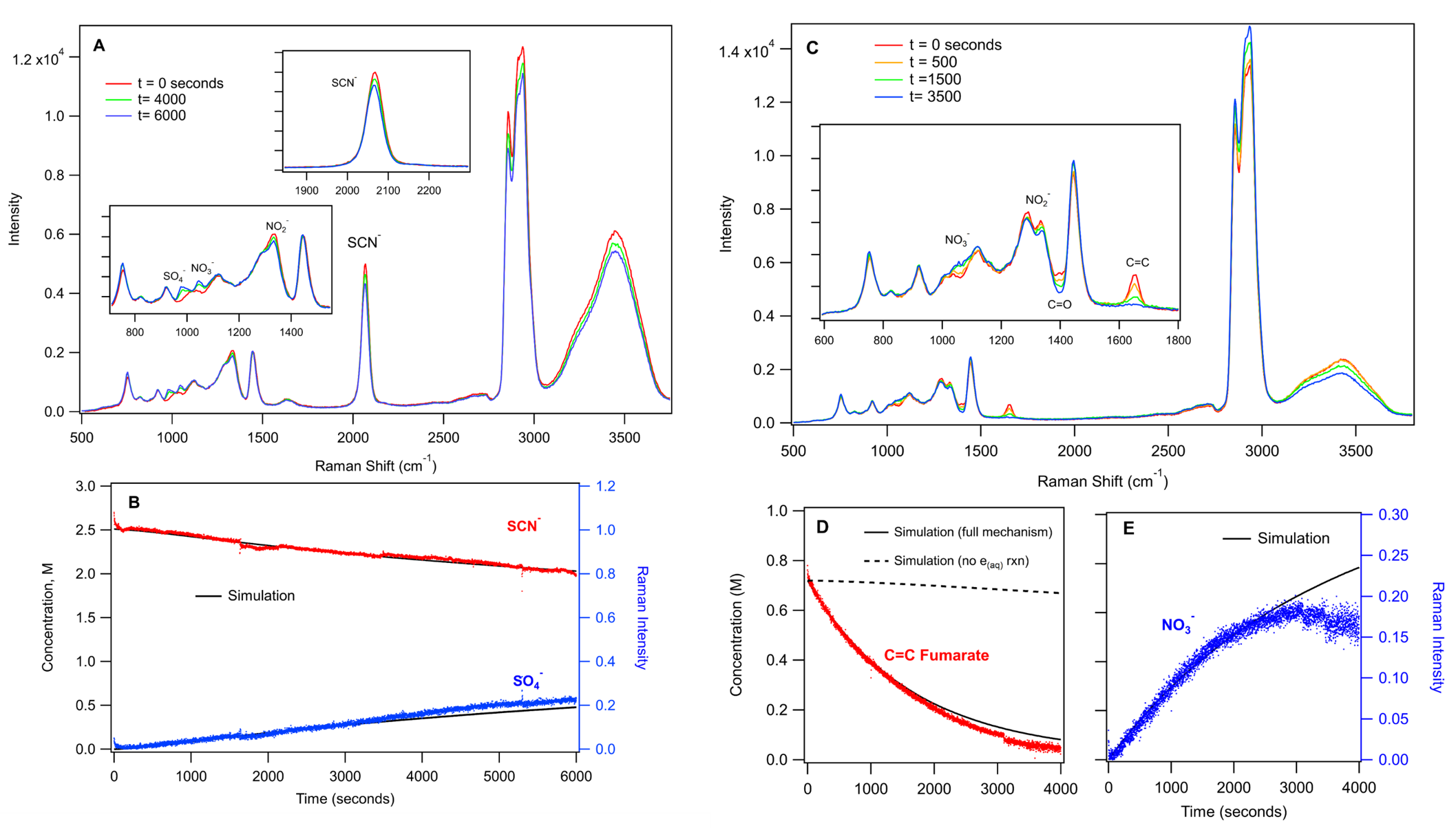}
\caption{Scavenger kinetics during nitrite photochemistry. (A) Raman spectra at select reaction times and (B) thiocyanate (SCN$^{-}$) decay and sulfate formation kinetics for $r \sim 32~\mu$m droplets. The initial droplet composition is: [SCN$^{-}$] = 2.51 M and [NO$_2^{-}$] = 2.5 M. (C) Raman spectra at select reaction times and (D) fumarate decay and (E) nitrate formation kinetics for $r \sim 33~\mu$m droplets. The initial droplet composition is: [Fumarate] = 0.72 M and [NO$_2^{-}$] = 2.38 M at pH = 6. Simulation results in panels (B), (D), and (E) are shown as solid black lines. In (B) and (E) the simulated sulfate and nitrate kinetics are compared with the relative Raman signal plotted on the right axis. In panel (D) fumarate kinetics are simulated with and without reactions with solvated electrons (i.e., with and without R24). The kinetics shown in (B), (D), and (E) are an average of 3 replicates.}
\label{fig:S9}
\end{figure}

The time dependent Raman spectra for fumarate are shown in Fig. S9C. The
decay kinetics (Fig. S9D) are monitored using fumarate's C=C band at
1651 cm\textsuperscript{-1}. Fumarate decays quickly and is accompanied
by a decrease in nitrite and the formation of nitrate (Fig. S9E). The
kinetic model replicates the decay kinetics of fumarate and formation of
nitrate as shown in Figs. S9D and S9E. Also shown in Fig. S9D are
kinetic model predictions when we turn off R24 (Table S3); the fumarate
+ \(e_{(aq)}^{-}\) reaction. Without R24, the model predicts kinetics
that are much too slow compared with experiment, indicating that the
fumarate lifetime is controlled mainly by its reactions with
\(e_{(aq)}^{-}\). Together these \({SCN}^{-}\)/fumarate scavenger
experiments verify the kinetic viability of our mechanism and
importantly provide evidence that both OH radicals and \(e_{(aq)}^{-}\)
are present; a distinctive feature of the photoionization mechanism.

As a final test, we use our kinetic model to predict previously measured
nitrite decay kinetics in microdroplets reported by Meroni \emph{et
al.}\cite{ref39} and Monroe \emph{et al.}\cite{ref40} Both studies use
aerosol optical tweezers (AOT) to measure droplet kinetics. Shown in
Figs. S10A and S10B are the results for nitrate formation as function of
size and initial {[}\({NO}_{2}^{-}\){]}. Using the estimated laser power
(see Fig. S1) at the sample position of the Biral AOT used by Meroni
\emph{et al.}\cite{ref39}, we are able to generate reasonable kinetic
predictions of their data as a function of droplet size and initial
nitrite concentration (Figs. S10A and S10B). The simulations are
sensitive to laser power and any differences between our Biral AOT
instrument and that used by Meroni \emph{et al.}\cite{ref39} would
naturally affect the overall agreement between model and experiment.
Nevertheless, the range of observed decay constants (2 x
10\textsuperscript{-4} s\textsuperscript{-1} \ensuremath{\leq}
\emph{k\textsubscript{obs}} \ensuremath{\leq} 1.1 x 10\textsuperscript{-3}
s\textsuperscript{-1}) reported by Meroni \emph{et al.}\cite{ref39} are
replicated in our full photoionization model and by Eq. (1) using
\(\epsilon \phi\) = 3.46 x 10\textsuperscript{-6}
M\textsuperscript{-1} cm\textsuperscript{-1} .

\begin{figure}[htbp]
\centering
\includegraphics[width=0.94\linewidth]{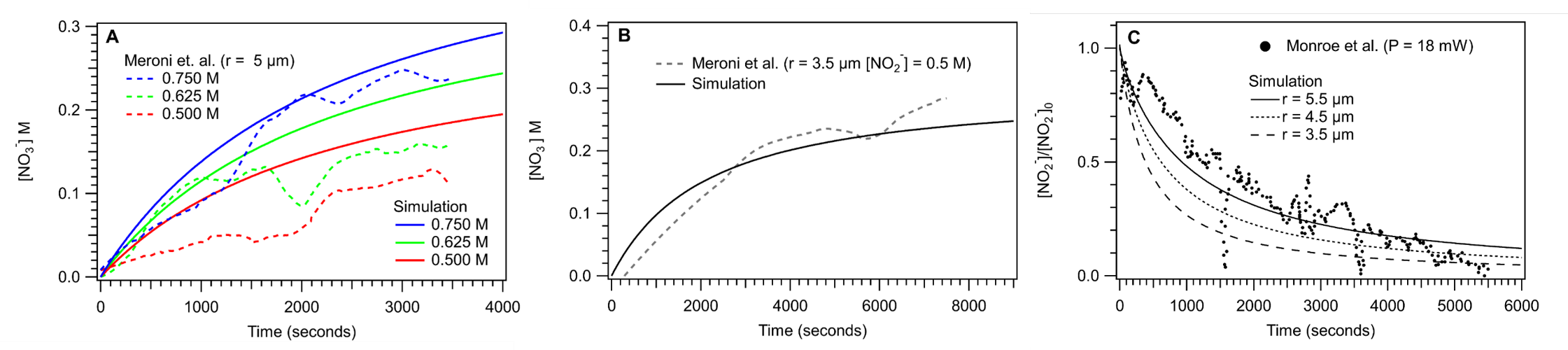}
\caption{(A) Nitrate formation kinetics as a function of [NO$_2^{-}$] for $r = 5~\mu$m. Dashed lines are data reported by Meroni \emph{et al.} Solid lines are results for kinetic simulations of the photoionization mechanism reported here. (B) Experimental (dashed lines, Meroni \emph{et al.}) and simulated (solid lines) nitrate formation kinetics for [NO$_2^{-}$] = 0.5 M and a droplet radius of 3.5 $\mu$m. (C) Normalized nitrite decay kinetics reported by Monroe \emph{et al.} Simulations are for $r = 3.5$, 4.5, and 5.5 $\mu$m and a laser power of 18 mW.}
\label{fig:S10}
\end{figure}

Shown in Fig. S10C is the normalized decay of {[}\({NO}_{2}^{-}\){]} vs.
time for a droplet comprised of 4 M nitrite.\cite{ref40} The simulations
are run for the droplet size range reported by Monroe \emph{et
al.}\cite{ref40} using the average corrected laser power on sample
reported in Ref. \cite{ref2} The experimental kinetics and associated
\emph{k\textsubscript{obs}} (\textasciitilde{} 10\textsuperscript{-3}
s\textsuperscript{-1}) are in reasonable agreement with predictions of
Eq. (1) (\(\epsilon \phi\) = 3.46 x 10\textsuperscript{-6}
M\textsuperscript{-1} cm\textsuperscript{-1}), and the full
photoionization kinetic model.

Although our photoionization mechanism provides a reasonable explanation
for the observations by Meroni \emph{et al.}\cite{ref39} and Monroe
\emph{et al.},\cite{ref40} the authors of both studies offer a rather
different interpretation of their data. Meroni \emph{et al.}\cite{ref39}
argue that the formation of nitrate from nitrite is due to the
spontaneous formation of OH radicals by a large intrinsic electric field
at the air-water interface. Alternatively, Monroe \emph{et
al.}\cite{ref40} attribute the decay of nitrite to autoxidation by
dissolved oxygen, with supporting evidence for this mechanism derived
from the slow/negligible reaction in a nitrogen atmosphere. While we
cannot definitively rule out autoxidation in the experiments conducted
by Monroe \emph{et al.},\cite{ref40} we would make the following
observations regarding the limitations of AOTs for measuring
photochemical reactions in droplets. In an AOT, it is difficult to
separately control the laser power required for trapping a droplet from
that required to drive photochemistry. This makes assessment of
photochemistry potentially difficult in an AOT. One approach, adopted by
others,\cite{ref6,ref41,ref42} is to measure the laser power dependences of
the reaction. Given these limitations, the photoionization mechanism
reported here, offers an alternative yet plausible explanation for these
previous results.

\begin{figure}[htbp]
\centering
\includegraphics[width=0.62\linewidth]{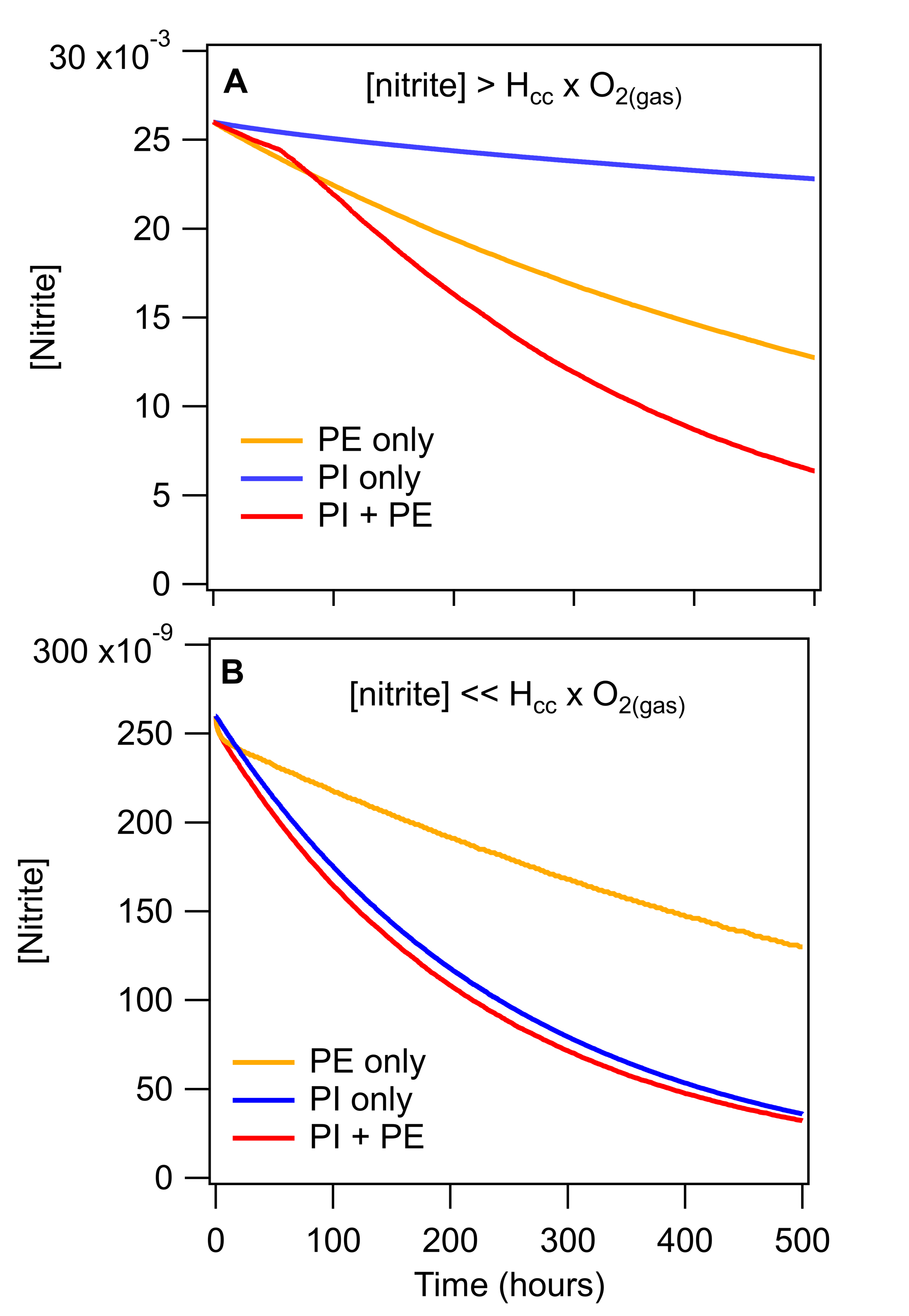}
\caption{Kinetic simulations of nitrite consumption by photoexcitation (PE) and photoionization (PI) at an initial $\left[\mathrm{NO}_{2}^{-}\right]$ of (A) $2.6 \times 10^{-2}$ M and (B) $2.6 \times 10^{-7}$ M.}
\label{fig:S11}
\end{figure}

\section*{S6. Kinetic simulations under dilute conditions}

Shown in Fig. S11 are simulations of the nitrite consumption by the
photoionization and photoexcitation mechanisms using the rate constants
(\emph{J}) determined in this work. The simulations are conducted using
Kinetiscope.\cite{ref43} The photoionization mechanism is shown in Tables
S3 and S4, while the photoexcitation mechanism is shown in Table S5.
Fig. S11A shows simulations for an initial
\(\left\lbrack {NO}_{2}^{-} \right\rbrack\) = 2.6 x
10\textsuperscript{-2} M and a dissolved {[}O\textsubscript{2}{]} = 2.8
x 10\textsuperscript{-4} M computed using the Henry's law constant
(\emph{H\textsubscript{cp}} = 0.00132 M/atm)\cite{ref37} for
O\textsubscript{2} in water. Under these conditions,
\(\left\lbrack {NO}_{2}^{-} \right\rbrack\) \textgreater{}
{[}O\textsubscript{2}{]} , where the main sink for the
\(e_{(aq)}^{\  -}\) is nitrite, photoexcitation consumes nitrite at a
rate that is 5x faster than photoionization. This contrasts with the
predictions for \(\left\lbrack {NO}_{2}^{-} \right\rbrack\) = 2.6 x
10\textsuperscript{-7} M (\emph{i.e.,}
\(\left\lbrack {NO}_{2}^{-} \right\rbrack\) $\ll $
{[}O\textsubscript{2}{]}), where photoionization consumes nitrite 3x
faster than photoexcitation. This shift in photochemical sinks with
concentrations originates from differences in the nitrite regeneration
pathways for photoexcitation (e.g., E5, E8, Table S5) and
photoionization (e.g., R8, R13, R19, etc. in Table S3).